\documentclass{article}

\usepackage{arxiv}

\author{
Thi-Thu-Tam Nguyen$^{1,2}$\\
$^{1}$Université Paris-Saclay, CentraleSupélec\\
  CNRS, Laboratoire des Signaux et Systèmes\\
  3 rue Joliot-Curie, 91190 Gif-sur-Yvette, France\\
  $^{2}$Pickup Services\\ 
  93400 Saint-Ouen, France\\	
   \And
Adnane Cabani$^{3}$\\
  $^{3}$ Université Rouen Normandie / ESIGELEC\\IRSEEM \\
76000 Rouen, France\\
	\And
Iyadh Cabani$^{2}$\\
  $^{2}$Pickup Services\\ 
  93400 Saint-Ouen, France\\	
   \And
Koen De Turck$^{4}$\\
$^{3}$ Department of Telecommunications and Information Processing\\
Ghent University, Sint-Pietersnieuwstraat 41\\
B-9000 Gent, Belgium\\
	\AND
Michel Kieffer$^{1}$\thanks{\textbf{Corresponding author: M. Kieffer, email: Michel.Kieffer@l2s.centralesupelec.fr}} \\
  $^{1}$Université Paris-Saclay, CentraleSupélec\\
  CNRS, Laboratoire des Signaux et Systèmes\\
  3 rue Joliot-Curie, 91190 Gif-sur-Yvette, France\\
}

\usepackage{graphics} 
\usepackage{epsfig} 
\usepackage{epstopdf} 
\usepackage{amsmath, adjustbox}
\usepackage{cuted}
\usepackage{amssymb}  
\usepackage[T1]{fontenc}

\usepackage[ruled,vlined]{algorithm2e}
\usepackage{xcolor}
\usepackage{multirow}
\usepackage{hyperref}
\usepackage{stfloats}

\newtheorem{prop}{Proposition}

\SetCommentSty{mycommfont}
\SetAlFnt{\small}

\title{\LARGE \bf
Forecasting the load of Parcel Pickup Points \\using a Markov Jump Process
}

\begin{document}

\maketitle
\thispagestyle{empty}
\pagestyle{empty}

\begin{abstract}
The growth of e-commerce has resulted in a surge in parcel deliveries, increasing transportation costs and pollution issues. Alternatives to home delivery have emerged, such as the delivery to so-called parcel pick-up points (PUPs), which eliminates delivery failure due to customers not being at home. Nevertheless, parcels reaching overloaded PUPs may need to be redirected to alternative PUPs, sometimes far from the chosen ones, which may generate customer dissatisfaction. 
Consequently, predicting the PUP load is critical for a PUP management company to infer the availability of PUPs for future orders and better balance parcel flows between PUPs. 

This paper proposes a new approach to forecasting the PUP load evolution using a Markov jump process that models the parcel life cycle. The latest known status of each parcel is considered to estimate its contribution to the future load of its target PUP. This approach can account for the variability of activity, the various parcel preparation delays by sellers, and the diversity of parcel carriers that may result in different delivery delays. Here, results are provided for predicting the load associated with parcels ordered from online retailers by customers (Business-to-Customer, B2C). The proposed approach is generic and can also be applied to other parcel flows to PUPs, such as second-hand products (Customer-to-Customer, C2C) sent via a PUP network.
\end{abstract}

\section{Introduction\label{sec:Introduction}}

The recent surge in e-commerce, mostly from online retailers to consumers (Business to Customer, B2C) \cite{rsf_b2c_2020}, has led to a substantial rise in parcel deliveries. This increase has implications for both transportation costs and environmental pollution. Furthermore, in case no one (neither customers, neighbors nor a concierge) is present to accept parcels, this results in failed delivery, necessitating multiple rescheduling attempts. In such situations, the distance covered by delivery services in increased substantially \cite{oliveira_analysis_2017}.

Alternative delivery services have been implemented, especially for
the last-mile delivery, \emph{i.e.}, the delivery process between
the last dispatch center of the carrier and the final customer. Customers can choose
to have their parcels delivered at a post office, or at a pick-up point
(PUP) close to their home or their workplace~\cite{morganti_final_2014}.
PUP Management Companies (PMCs) offer two types of PUPs~\cite{weltevreden_b2c_2008}:
\emph{i}) automatic parcel lockers (APL)~\cite{iwan_analysis_2016},
usually installed within train stations, supermarkets, or \emph{ii})
local shops (such as food stores or corner shops). This delivery service
presents multiple advantages, including a wider range of opening hours
compared to that of post offices for customers to pick up their parcels, 
as well as additional customer visits and income for local shops serving as
PUPs~\cite{weltevreden_b2c_2008}. Finally, it is also a sustainable
solution for reducing transportation fees and limiting delivery failures
compared to traditional home delivery~\cite{zhou_understanding_2020}.
Customers can also drop off parcels at these PUPs in case of product
returns or when they sell new or used products to other customers
(C2C service).

Managing a PUP network comes with several challenges.
When the chosen PUP is overloaded, customers may find their parcels
delivered at a different PUP, sometimes far from their intended
pick-up location, leading to customer dissatisfaction. Local shops 
serving as PUPs may sometimes receive too many parcels for 
their storage capacity. The parcel
management activity may then be detrimental to their primary activity,
especially when some parcels have not been accepted and have been
re-routed to an alternative PUP. Some other PUPs do not handle enough
parcels to benefit from the PUP activity. These are the two main reasons
for contract cancellation betwen PUPs and PMCs.

Consequently, the PMCs have to monitor and control the load of each PUP carefully
in order to better balance parcel loads among neighboring ones.
PUPs likely to be overloaded in the coming days will not
be offered to customers during the ordering process. To achieve this,
predicting the load of a given PUP several days in advance is essential
for PMCs to manage their PUP network more effectively.

Our previous paper~\cite{nguyen_load_2022} describes forecasting
approaches for the load associated with the Business-to-Customer (B2C)
process by considering the load evolution as a time series. This approach 
makes it difficult to account for the variability of sellers and of carriers,
which may introduce a large diversity of parcel preparation and delivery delays.
In this work, we model the latency between statuses 
for each parcel rather than estimating the 
number of parcels in each status, as done in~\cite{nguyen_load_2022}.
We assume that each parcel runs through a non-stationary 
Markov chain where the states represent the different statuses the parcel can take on.  
By doing so, we can deduce
the probability distribution of the load at each instant, which is a
critical information for PMC decisions regarding the availability
of PUPs for future orders. 
The temporal variability of the activity (peaks during sales or before Christmas), as well 
as the diversity of sellers and carriers are then easier to take into account. 
Moreover, the adopted approach is generic and allows for different status flowcharts on the 
part of the PMCs, and provides a much larger modeling flexibility compared to \cite{nguyen_load_2022}.
We subsequently apply this approach to a specific scenario.

The remainder of the paper is organized as follows. Section~\ref{sec:Related-works}
presents some related works. Section~\ref{sec:Problem-description}
details and formalizes the PUP load forecasting problem. Section~\ref{sec:Pred_Lt}
describes the load forecasting approach. The considered prediction
approach is compared with alternative ones in terms of prediction
accuracy in Section~\ref{sec:Results}. Conclusions and perspectives
are provided in Section~\ref{sec:Conclusion}. 

Table~\ref{tab:Main-notations} introduces the main notations used in this paper.

\begin{table}[ht]
\begin{centering}
\begin{tabular}{|c|p{14cm}|}
\hline 
Notation  & Variables\tabularnewline
\hline 
$\tau$  & Parcel identifier\tabularnewline
\hline 
$\rho$  & PUP identifier\tabularnewline
\hline 
$T$  & Time sampling period\tabularnewline
\hline 
$k$  & Time interval index\tabularnewline
\hline
$H_{0}$ & Necessary time to prepare the order for expedition\tabularnewline
\hline
$H_{n}$ & Holding time for a transition between two intermediate state\tabularnewline
\hline
$\mathcal{R}$ & The set of retailers\tabularnewline
\hline
$\Gamma$ & The set of available carriers\tabularnewline
\hline 
$\mathcal{S}$ & The set of status\tabularnewline
\hline
\multicolumn{2}{|l|}{At time $k$}\tabularnewline
\hline 
$w\left(k\right)$  & the day of the week of the time interval $k$\tabularnewline
\hline 
$h$$\left(k\right)$  & the hour of the day of time $k$\tabularnewline

\hline 
$\mathcal{P}_{k}$  & set of parcel indexes with information available with any PUP as target\tabularnewline
\hline 
$\mathcal{P}_{k}^{\rho}$  &set of parcel indexes with information available with PUP $\rho$ as target\tabularnewline
\hline
$\mathbb{V}_{k}$ & set of virtual parcel indexes (since not known at time $k$) \tabularnewline
\hline 
$\Lambda_{\tau}\left(k\right)$  & variable indicating whether $\tau$ contributes to the load of PUP
$\rho$\tabularnewline
\hline 
$L\left(k\right)$ & number of parcels in target PUP\tabularnewline
\hline 
$O\left(k\right)$  & number of orders that will be confirmed at time $k$\tabularnewline
\hline 
\multicolumn{2}{|l|}{For parcel $\tau$}\tabularnewline
\hline
$R(\tau)$ & bought from the retailer $R$\tabularnewline
\hline
$C(\tau)$ & carrier in charge of the delivery\tabularnewline
\hline 
$S_{k}\left(\tau\right)$  & current status during time interval $k$\tabularnewline
\hline 
$T_{S_{k}}\left(\tau\right)$  & first time interval at which status has switched to $S_{k}$\tabularnewline
\hline
$H_{n}$ & holding time in status $n$\tabularnewline
\hline 
\multicolumn{2}{|l|}{Sets of identifiers of parcels expected to contribute to target PUP
load at time $k+j$}\tabularnewline
\hline 
$\mathcal{L}_{N}\left(k+j\mid k\right)$ & parcels known to be delivered (status $N-1$) before time $k$, and expected to be still waiting to be picked-up at time $k+j$\tabularnewline
\hline
$\mathcal{L}_{n}\left(k+j\mid k\right)$ & parcels with known status $n-1$ before time $k$, and which expected to be delivered before time $k+j$\tabularnewline
\hline
$\mathcal{L}_{0}\left(k+j\mid k\right)$ & parcels containing products not yet ordered at time $k$, but which are expected to be waiting to
be picked-up at time $k+j$\tabularnewline
\hline 
\end{tabular}
\par\end{centering}
\caption{Main notations\label{tab:Main-notations}}
\end{table}

\section{Related work\label{sec:Related-works}}

The development of last-mile delivery raises multiple difficulties~\cite{cardenas_city_2017,bosona_urban_2020},
such as customer desires for shorter delivery delays, seasonal peaks
of parcels \cite{allen_understanding_2018}, optimal deployment of
PUP locations \cite{wang_robust_2022}, \emph{etc}.

While the limited-capacity issue of APLs has been identified in \cite{gevaers_characteristics_2009}, the load of PUPs has not been studied in detail, although this aspect needs to be controlled by the PMCs in order to limit delivery failures.
The PUP parcel load prediction consists of the evaluation of the parcel
preparation and transportation delays once a product has been ordered,
and the prediction of the pick-up delay by the e-customer once the parcel has been delivered to the PUP. 

Several works have proposed prediction approaches for e-commerce activity. In \cite{abbasimehr_optimized_2020}, statistical and computational intelligence methods are put at work using demand data of a furniture company. A multi-layer LSTM model is employed and compared to alternative techniques. A data mining approach is proposed \cite{zhang_comparative_2023} for online clothing sales forecasting. Seasons, sales, and holidays are identified as essential factors of demand in \cite{kurata_optimal_2007}.
The time-series forecasting library Prophet \cite{taylor_forecasting_2017}
and support vector regression models \cite{cristianini_introduction_2000}
are combined in \cite{guo_hybrid_2021} to forecast time series demand
in the manufacturing industry accounting for seasonality.

Regarding the delivery delay estimation, \cite{salari_real-time_2020} proposes a real-time forecasting approach of the delivery time based on relevant operational features, including the time of order, the distribution center that will be used, the order, and the user.
Concerning the pick-up delay, a statistical analysis in \cite{parcelmonitor_dwell_2021} shows
that more than $60$~\% of the parcels are collected less than 24
hours after delivery, and about $75$~\% are collected in less than
48 hours, with considerable regional variations.

While the prediction of each of these aspects (amount of sales, delivery
delays, and pick-up delays) have been studied separately, there is
a gap in the literature as far as a combined analysis is concerned
which forms the subject of this paper. Similar multi-faceted forecasting
problems can also be identified for the load evaluation of an intermediate
warehouse, where the duration between the delivery and the pick-up
process has to be evaluated~\cite{brajon_comment_2016}.

In this paper, we extensively use Markov models~\cite{stewart_elementary_2009},
frequently employed to describe systems with discrete state
transitions, assuming (for first-order models) that the transition
probabilities only depend on the current state of the system. Considering
the life cycle of parcels, transition probabilities from a given state
evolve with time. In our study,
factors determining the state transition probabilities of a parcel
at each time instant are primarily the current state, the day of the
week, and the hour of the last transition. These characteristics are
closer to a non-homogeneous (or non-stationary) Markov model, as described
in~\cite{vassiliou_non-homogeneous_2021}, where the transition probabilities 
depend not only on the current state but also on the current time instant. Other particularities of our process are the transitions
to one direction only, see Figure~\ref{fig:transition-states}, and
the maximum sojourn time of each parcel after delivery. 
Therefore we consider a non-stationary Markov model taking into 
account all these characteristics, which will be developed in Section~\ref{sec:Problem-description}.

\section{Problem description\label{sec:Problem-description}}

This section models the life-cycle of a parcel starting from its order
confirmation on the website of an online retailer (classical or second-hand)
and ending with its pickup by a customer in the target PUP. The
transitions between the various parcel states are described by a Markov
jump process. 
Finally, the PUP load forecasting problem is formalized.

\subsection{Model of the life-cycle of a parcel}

Time is sampled with a period of $T$ (typically one hour). Let $k\in\mathbb{N}$
be the index of the time interval $\left[kT,\left(k+1\right)T\right[$.
The day of the week of the time interval $k$ is $w\left(k\right)\in\left[1,7\right]$,
from Monday ($w\left(k\right)=1$) to Sunday ($w\left(k\right)=7$).
The hour of the day at which the interval $\left[kT,\left(k+1\right)T\right[$
starts is $h\left(k\right)\in\left[0,24\right[$.

Consider a parcel with index $\tau$ bought from an online retailer
$R\left(\tau\right)\in\mathcal{R},$ where $\mathcal{R}$ is the
set of retailers. A carrier $C\left(\tau\right)\in\Gamma$ is in charge
of the delivery of the parcel to some target PUP $\rho$, where $\Gamma$
is the set of available carriers. The evolution of the state of parcel
$\tau$ is represented by the sequence of random pairs $\left\{ \left(S_{k}\left(\tau\right),T_{S_{k}}\left(\tau\right)\right)\right\} _{k\in\mathbb{N}}$,
where $S_{k}\left(\tau\right)\in\mathcal{S}=\left\{ 0\dots N\right\} $
indicates the status of the parcel $\tau$ in the time interval $k$
and 
\begin{equation}
T_{S_{k}}\left(\tau\right)=\min\left\{ \ell\,|\,S_{\ell}\left(\tau\right)=S_{k}\left(\tau\right)\right\} \label{eq:defineT_sk}
\end{equation}
represents the index of the time interval at which the parcel has
switched to status $S_{k}\left(\tau\right)$\footnote{The argument $\tau$ will be sometimes omitted to
lighten notations.}. If $S_{k+1}=S_{k}$ then $T_{S_{k+1}}=T_{S_{k}}$. If $S_{k+1}\neq S_{k}$
then $T_{S_{k+1}}=k+1$. Consequently, there is a minimum delay of
$T$ between consecutive transitions. When the order is confirmed,
the parcel is in status $S=0$. Parcels with status $S=N-1$ have
been delivered to the PUP. Finally, the status $S=N$ corresponds
to a parcel picked-up by the customer or returned to the retailer.
All intermediate statuses between $1$ and $N-2$ correspond to the
product wrap-up, collection at the retailer warehouse,
and processing at the intermediate logistic platforms of the
carrier.

\begin{figure}[ht]
\centering\includegraphics[width=0.4\textwidth]{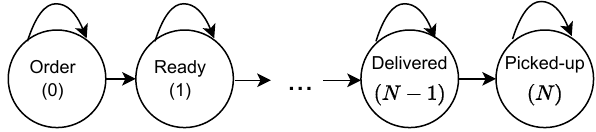}\caption{Statuses and possible transitions between statuses for a parcel; From
time $k$ to $k+1$, the parcel can stay in status $n$ or switch
to the status $n+1$; When the status $N$ is reached (picked-up parcel),
parcels stay in this status and do not contribute to the PUP load
anymore. \label{fig:transition-states}}
\end{figure}

The transitions between these statuses are presented in Figure~\ref{fig:transition-states}.
We assume that the sequence of states of the parcel $\left\{ \left(S_{k},T_{S_{k}}\right)\right\} _{k\in\mathbb{N}}$
satisfies the Markov property 
\begin{align*}
\mathbb{P} &\left(\left(s_{k+1},t_{s_{k+1}}\right)\mid\left(s_{k},t_{s_{k}}\right),\dots,\left(s_{1},t_{s_{1}}\right)\right)\nonumber\\
 & =\mathbb{P}\left(\left(s_{k+1},t_{s_{k+1}}\right)|\left(s_{k},t_{s_{k}}\right)\right).
\end{align*}

For all $n\in\mathcal{S}$, $n<N$, the holding time of parcel $\tau$
in status $n$ is 
\begin{equation}
H_{n}=T_{n+1}-T_{n}.\label{eq:H_n}
\end{equation}
\vspace{-0.8cm}
 
The probability for parcel $\tau$ to switch to state $\left(n+1,t_{n+1}\right)$,
$t_{n+1}-t_{n}$ time slots after having entered state $\left(n,t_{n}\right)$
at time $t_{n}<t_{n+1}$ depends on several parameters, such as the
time $T_{n}=t_{n}$, the online retailer $R$, the carrier $C$, and
the target PUP~$\rho$. In the most general case, for $0\leqslant n<N$,
one has 
\begin{align}
\mathbb{P} & \left(T_{n+1}=t_{n+1}\mid T_{n}=t_{n},R=r,C=c,\rho\right)\nonumber \\
 & =\mathbb{P}\left(H_{n}=t_{n+1}-t_{n}\mid T_{n}=t_{n},R=r,C=c,\rho\right)\nonumber \\
 & =f_{n}\left(t_{n+1}-t_{n}\mid t_{n},r,c,\rho\right).\label{eq:def_f_n}
\end{align}
Nevertheless, for a given status $n$, the transition probability
in (\ref{eq:def_f_n}) may depend only on a reduced subset of parameters.

For example, if status $S=1$ corresponds to a parcel wrapped up and
ready to be taken over at the retailer warehouse, the holding time
$H_{0}$ represents the time necessary to prepare the order for expedition.
This time depends only on the retailer $R=r$ and on the order confirmation
time $T_{0}=t_{0}$. One may further assume that $H_{0}$ only depends
on the day of the week $w\left(t_{0}\right)$ and the hour of the
day $h\left(t_{0}\right)$ of the order confirmation. Once a parcel
has been taken over by a carrier $C=c$, the following holding times
do not depend any more on the retailer. Consequently, for a transition
between two intermediate states $\left(n,t_{n}\right)$ to $\left(n+1,t_{n+1}\right)$
with $n>1$ and $n+1<N$ during the parcel transportation,
the holding time $H_{n}$ depends mainly on the carrier $C=c$, as
well as the day of the week $w\left(t_{n}\right)$ and the hour of
the day $h\left(t_{n}\right)$ of the status $n$. The time of delivery
to the PUP $T_{N-1}=t_{N-1}$ depends on the carrier $C=c$, the day
of the week $w\left(t_{N-2}\right)$ and the hour of the day $h\left(t_{N-2}\right)$
of the status $N-2$, and on the PUP $\rho$ (parcel are delivered
to PUPs every day, except Sundays over a relatively short time interval of the day, depending
on the tour of the carrier).

Accounting for fewer parameters in the expression of $f_{n}$ facilitates
its estimation from historical data. In what follows, the dependency
of $f_{n}$ in $r$, $c$, or $\rho$ is omitted to lighten notations.

\subsection{Problem formulation\label{subsec:Problem-formulation}}

A parcel $\tau$ contributes to the load of PUP~$\rho$ at time $k$
if $S_{k}\left(\tau\right)=N-1$, \emph{i.e.}, if $T_{N-1}\left(\tau\right)\leqslant k$ (it has been delivered to the PUP before time $k$)
and $T_{N}\left(\tau\right)>k$ (it has not been picked-up at time $k$). At time $k$, the total number of
parcels stored in PUP~$\rho$ is then 
\begin{align}
L\left(k\right) & =\text{Card}\left(\left\{ \tau\;\text{such that}\;S_{k}\left(\tau\right)=N-1\right\} \right)\label{eq:L_t} \\
 & =\text{Card}\left(\left\{ \tau\;\text{such that}\;T_{N-1}\left(\tau\right)\leqslant k,T_{N}\left(\tau\right)>k\right\} \right).\nonumber
\end{align}

The aim of this paper is to build estimators $L\left(k+j|k\right)$
of the load $L\left(k+j\right)$ at time $k+j$, $j=1\dots j_{\text{max}}$,
over a prediction horizon of up to $j_{\text{max}}$ time intervals,
using only the information available at time $k$ and related to the
parcels which have been or will be delivered to the considered PUP.

\section{Prediction of $L$\label{sec:Pred_Lt}}

The evaluation at time $k$ of the predicted load $L\left(k+j\mid k\right)$
requires the evaluation of the pmf of $T_{N-1}\left(\tau\right)$
and $T_{N}\left(\tau\right)$ for each parcel $\tau$ potentially contributing
to the load of PUP~$\rho$. The expressions of these pmfs 
depend on the status of parcel $\tau$
known at time $k$. Section~\ref{subsec:Sets-of-parcels} introduces
a partition of the set of parcels contributing to $L\left(k+j\mid k\right)$
as a function of their status at time $k$. Then Section~\ref{subsec:set_contribution}
describes the way the pmf of $T_{N-1}\left(\tau\right)$ and $T_{N}\left(\tau\right)$
are obtained as a function of the status of parcel $\tau$. Some products
have not been ordered at time $k$ but may contribute to $L\left(k+j\right)$.
Section~\ref{subsec:Prediction_DO} introduces a model of the number
of future orders with PUP $\rho$ as target to account for their contribution
to $L\left(k+j\right)$. Finally, Section~\ref{subsec:Load-prediction-algorithm}
summarizes the load prediction algorithm.

\subsection{Sets of parcels contributing to the load of the PUP\label{subsec:Sets-of-parcels}}

\begin{figure}[ht]
\centering\includegraphics[width=0.45\columnwidth]{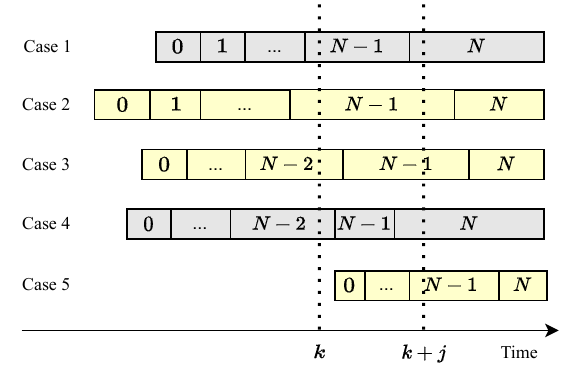}

\caption{Illustration of parcels contributing to the load $L$$\left(k+j\right)$
for a prediction performed at time $k$. One has to account for parcels
delivered before time $k$ (cases 1 and 2) as well as parcels taken
over by a carrier before time $k$ and not yet delivered (cases 3
and 4). Only parcels expected to be picked up after time $k+j$ will
actually contribute to the load at time $k+j$ (cases 2 and 3). Products
that are expected to be ordered between time $k$ and time $k+j$
(case 5) may also contribute to the load at time $k+j$.\label{fig:Decomposition-Lt}}
\end{figure}

Using the information available at time $k$, the set $\mathcal{L}\left(k+j\mid k\right)$
of parcels that contribute to $L\left(k+j\right)$ is partitioned
into several subsets. The set 
\begin{align*}
\mathcal{L}_{N}\!\left(k+j\mid k\!\right) & =\!\left\{ \tau\mid T_{N-1}\!\left(\tau\!\right)\leqslant k,T_{N}\!\left(\tau\!\right)>k+j\!\right\} 
\end{align*}
contains the indexes of parcels known to be delivered (Status $N-1$)
to the considered PUP before time $k$ and expected to be still waiting
to be picked-up at time $k+j$. The set 
\begin{align*}
\mathcal{L}_{n}\left(k+j\mid k\right) & =\left\{ \tau\mid T_{n-1}\left(\tau\right)\leqslant k,T_{n}\left(\tau\right)>k,
 T_{N-1}\left(\tau\right)\leqslant k+j,T_{N}\left(\tau\right)>k+j\right\} 
\end{align*}
contains the indexes of parcels with known status $n-1$ before
time $k$, with $0<n<N$, for which $T_{n}\left(\tau\right)>k$ is unknown at time
$k$, and which are expected to be delivered to the PUP before time
$k+j$ and to be still waiting to be picked-up at time $k+j$. Finally,
the set
\[
\mathcal{L}_{0}\!\left(k+j\mid k\!\right)\!=\!\left\{ \tau\mid T_{0}\!\left(\tau\!\right)\!>k,T_{N-1}\!\left(\tau\!\right)\!\leqslant k+j,T_{N}\!\left(\tau\!\right)\!>k+j\!\right\}
\]
contains the indexes of parcels containing products not
yet ordered at time $k$, but which are expected to be waiting to
be picked-up in the target PUP~$\rho$ at time $k+j$. Again, these
parcels need to be delivered before time $k+j$.

Consequently, 
\begin{equation}
\mathcal{L}\left(k+j\mid k\right)=\bigcup_{n=0}^{N}\mathcal{L}_{n}\left(k+j\mid k\right).\label{eq:PartitionofL}
\end{equation}
Figure~\ref{fig:Decomposition-Lt} illustrates parcels with different
status at time $k$ and the evolution of their status with time. Only
a subset of these parcels contributes to the load of the PUP~$\rho$
at time $k+j$.

\subsection{Contribution to the load for each set of parcels}

\label{subsec:set_contribution}

Let $\Lambda_{\tau}\left(k\right)$ be the random variable describing
the contribution of each parcel $\tau$ to the load of the PUP~$\rho$
at time $k$, \emph{i.e.}, $\Lambda_{\tau}\left(k\right)=1$ if the
parcel waits to be picked-up at time $k$ and $\Lambda_{\tau}\left(k\right)=0$
else. Let $\mathcal{P}_{k}^{\rho}$ be the set of parcels with target
PUP $\rho$ for which information is available at time $k$. Consequently
\begin{equation}
L\left(k\right)=\textstyle \sum_{\tau\in\mathcal{P}_{k}^{\rho}}\Lambda_{\tau}\left(k\right).\label{eq:L_{k}}
\end{equation}

At time $k$, to evaluate the predicted load $L\left(k+j\mid k\right)$
at time $k+j$, one considers the partition of parcels introduced
in \eqref{eq:PartitionofL} to get 
\begin{align}
L\left(k+j\mid k\right)={\textstyle \sum_{n=0}^{N}L_{n}\left(k+j\mid k\right),\label{eq:sum_{L}}}
\end{align}
where $L_{n}\left(k+j\mid k\right)$ is the cardinal number of the
set $\mathcal{L}_{n}\left(k+j\mid k\right)$. The random variables
$L_{n}\left(k+j\mid k\right)$ are assumed to be independent. Consequently,
the pmf of $L\left(k+j\mid k\right)$ is obtained as the convolution
of the pmfs of all $L_{n}\left(k+j\mid k\right)$, $n=0,\dots,N$.

In this section, considering the status of a parcel $\tau$ at time
$k$, one evaluates its probability to belong to $\mathcal{L}\left(k+j\mid k\right)$,
\emph{i.e.}, to contribute to the load of the considered PUP at time
$k+j$.
\begin{prop}
\label{prop:Ld}At time $k$, consider a parcel $\tau$ with current
status $S_{k}=N-1$, \emph{i.e.}, that has been delivered at time
$t_{N-1}\leqslant k$ and that has not yet been picked up time $k$.
The pickup time $T_{N}$ depends on the delivery time $T_{N-1}$,
and implicitly on the opening hours of the PUP~$\rho$. The probability
that $\tau\in\mathcal{L}_{N}\left(k+j\mid k\right)$, \emph{i.e.},
that it is still in the PUP at time $k+j$, $j\geqslant1$ is
\begin{align}
\mathbb{P}\left(\Lambda_{\tau}\left(k+j\right)=1\mid T_{N-1}=t_{N-1},T_{N}>k\right)
 &=\mathbb{P}\left(T_{N}>k+j\mid T_{N-1}=t_{N-1},T_{N}>k\right)\nonumber \\
 & =\frac{1-\sum_{t_{N}=t_{N-1}+1}^{k+j}f_{N-1}\left(t_{N}-t_{N-1}\mid t_{N-1}\right)}{1-\sum_{t_{N}=t_{N-1}+1}^{k}f_{N-1}\left(t_{N}-t_{N-1}\mid t_{N-1}\right)}.\label{eq:L_d_final}
\end{align}
\end{prop}
The proof of Proposition~\ref{prop:Ld} is provided in Appendix~\ref{subsec:Proof_Prop_Ld}.

\begin{prop}
\label{prop:Lc} At time $k$, consider a parcel $\tau$ with current
status $S_{k}=N-2$ and carrier $C\left(\tau\right)=c$. This parcel
is still in transit to the target PUP at time $k$. The delivery time
$T_{N-1}$ depends on the carrier $c$ and on the time $T_{N-2}$
(of collection, \emph{e.g.}, at the dispatch center closest to the
PUP). The probability that $\tau\in\mathcal{L}_{N-1}\left(k+j\mid k\right)$,
\emph{i.e.}, that it is delivered before time $k+j$ and still in
the PUP at time $k+j$, $j\geqslant1$ is 

\begin{adjustbox}{max width=\textwidth}%
\noindent\parbox[c]{1\linewidth}{%
\begin{align}
 \mathbb{P}&\left(\Lambda_{\tau}\left(k+j\right)=1\mid T_{N-2}=t_{N-2},T_{N-1}>k,c\right)
  =\mathbb{P}\!\left(T_{N-1}\!\leqslant k+j,T_{N}\!>k+j|T_{N-2}\!=t_{N-2},T_{N-1}\!>k,c\right),\nonumber \\
 & =\frac{\displaystyle\sum_{t_{N-1}=k+1}^{k+j}f_{N-2}\left(t_{N-1}-t_{N-2}\mid t_{N-2}\right)}{1-\displaystyle\sum_{t_{N-1}=t_{N-2}+1}^{k}f_{N-2}\left(t_{N-1}-t_{N-2}\mid t_{N-2}\right)}
 \cdot\left(1-\frac{\displaystyle\sum_{t_{N-1}=k+1}^{k+j-1}\sum_{t_{N}=t_{N-1}+1}^{k+j}f_{N-1}\left(t_{N}-t_{N-1}\mid t_{N-1}\right)f_{N-2}\left(t_{N-1}-t_{N-2}\mid t_{N-2}\right)}{\displaystyle\sum_{t_{N-1}=k+1}^{k+j}f_{N-2}\left(t_{N-1}-t_{N-2}\mid t_{N-2}\right)}\right).\label{eq:L_o_final}
\end{align}}
\end{adjustbox}
\end{prop}

The proof of Proposition~\ref{prop:Ld} is provided in Appendix~\ref{subsec:Proof_Prop_Lc}.
In \eqref{eq:L_o_final}, the dependency in $c$ of $f_{N-2}$ has
been omitted.
\begin{prop}
\label{prop:Ln} At time $k$, consider a parcel $\tau$ with current
status $S_{k}=n$, $0<n<N-2$, reached at time $t_{n}$. The parcel
is shipped by carrier $C\left(\tau\right)=c$ to the target PUP $\rho$.
The probability that $\tau\in\mathcal{L}_{n+1}\left(k+j\mid k\right)$,
\emph{i.e.}, that it is delivered before time $k+j$ and still in
the PUP at time $k+j$, $j\geqslant1$ is 

\begin{adjustbox}{max width=\textwidth}%
\noindent\parbox[c]{1\linewidth}{%
\begin{align}
 \mathbb{P}&\left(\Lambda_{\tau}\left(k+j\right)=1\mid T_{n}=t_{n},T_{n+1}>k,r,c\right)
 =\mathbb{P}\!\left(T_{N-1}\!\leqslant k+j,T_{N}\!>k+j\!\mid T_{n}=t_{n},T_{n+1}\!>k,r,c\right) \nonumber \\
 & =\frac{\sum_{t_{N-1}=k+1}^{k+j}g_{n}^{N-1}\left(t_{n},t_{N-1}\right)}{1-\sum_{t_{n+1}=t_{n}+1}^{k}f_{n}\left(t_{n+1}-t_{n}\mid t_{n}\right)}
  \cdot\left(1-\frac{\displaystyle\sum_{t_{N}=k+1}^{k+j}\displaystyle\sum_{t_{N-1}=k+1}^{t_{N}}\sum_{t_{n+1}=k+1}^{t_{N-1}}f_{N-1}\left(t_{N}-t_{N-1}\mid t_{N-1}\right)g_{n+1}^{N-1}\left(t_{n+1},t_{N-1}\right)f_{n}\left(t_{n+1}-t_{n}\mid t_{n}\right)}{\displaystyle\sum_{t_{N-1}=k+1}^{k+j}\sum_{t_{n+1}=k+1}^{t_{N-1}}g_{n+1}^{N-1}\left(t_{n+1},t_{N-1}\right)f_{n}\left(t_{n+1}-t_{n}\mid t_{n}\right)}\right),\label{eq:exp_L_n}
\end{align}}
\end{adjustbox}

\noindent The function 
\begin{align}
g_{n}^{N-1}\left(t_{n},t_{N-1}\right) & =\sum_{t_{n+1}=k+1}^{t_{N-1}}\sum_{t_{n+2}=t_{n+1}+1}^{t_{N-1}}\dots\sum_{t_{N-2}=t_{N-3}+1}^{t_{N-1}}
 f_{N-2}\left(t_{N-1}-t_{N-2}\mid t_{N-2}\right)\dots\nonumber \\
 & f_{n+1}\left(t_{n+2}-t_{n+1}\mid t_{n+1}\right)f_{n}\left(t_{n+1}-t_{n}\mid t_{n}\right).\label{eq:gnN}
\end{align}
in \eqref{eq:gnN} represents 
the probability for a parcel to switch from status $n$ reached
at time $t_{n}$ to status $N-1$ at time $t_{N-1}$.
\end{prop}

The proof of Proposition~\ref{prop:Ln} is provided in Appendix~\ref{subsec:Proof_Prop_Ln}. 

In \eqref{eq:exp_L_n} and \eqref{eq:gnN}, the dependency of $f_{n}$
with the carrier $c$ has been omitted. Moreover, in \eqref{eq:gnN},
all sums have been written with up to $t_{N-1}$ to lighten notations.
Nevertheless, as there is at least one time interval $T$ between
consecutive transitions, one has $f_{n+1}\left(0\mid t_{n+1},c\right)=0$,
and many terms in \eqref{eq:gnN} will vanish.

\begin{prop}
\label{prop:L0} At time $k$, consider a parcel $\tau$ that will
be ordered at time $T_{0}\left(\tau\right)=t_{0}$ with $k<t_{0}<k+j$
with target PUP $\rho$. The parcel preparation duration $H_{0}\left(\tau\right)=T_{1}\left(\tau\right)-T_{0}\left(\tau\right)$
depends on the retailer $r$, and the delivery time depends on the
carrier $c$. The probability that $\tau\in\mathcal{L}_{\text{0}}\left(k+j\mid k\right)$,
\emph{i.e.}, that it is delivered before time $k+j$ and still in
the PUP at time $k+j$, $j\geqslant1$ is 

\begin{adjustbox}{max width=\textwidth}%
\noindent\parbox[c]{1\linewidth}{%
\begin{align}
 \mathbb{P}&\left(\Lambda_{\tau}\left(k+j\right)=1\mid T_{0}=t_{0},r,c\right)=\mathbb{P}\left(T_{N-1}\leqslant k+j,T_{N}>k+j\mid T_{0}=t_{0},r,c\right)\nonumber \\
 & =\sum_{t_{N-1}=k+1}^{k+j}\sum_{t_{0}=k+1}^{t_{N-1}}g_{0}^{N-1}\left(t_{0},t_{N-1}\right)
 \cdot\left(1-\frac{\displaystyle\sum_{t_{0}=k+1}^{k+j-\left(N-1\right)}\sum_{t_{N-1}=t_{0}+N-1}^{k+j-1}\sum_{t_{N}=t_{N-1}+1}^{k+j}f_{N-1}\left(t_{N}-t_{N-1}\mid t_{N-1}\right)g_{0}^{N-1}\left(t_{0},t_{N-1}\right)}{\displaystyle\sum_{t_{0}=k+1}^{k+j-\left(N-1\right)}\sum_{t_{N-1}=t_{0}+N-1}^{k+j}g_{0}^{N-1}\left(t_{0},t_{N-1}\right)}\right).\label{eq:L0expanded}
\end{align}}
\end{adjustbox}
\end{prop}

The proof of Proposition~\ref{prop:L0} can be found in Appendix~\ref{App:Proof:L0}.
In \eqref{eq:L0expanded}, the dependency of \eqref{eq:L0expanded}
in $r$ and $c$ has also been omitted.

The number of parcels that may contribute to $\mathcal{L}_{\text{0}}\left(k+j\mid k\right)$
is not known at time $k$, contrary to parcels that may contribute
to $\mathcal{L}_{n}\left(k+j\mid k\right)$, $0<n\leqslant N-1$ for
which more information is available (their order is at least confirmed).
To address this issue, we choose to model the number of orders $O\left(k+i \mid k\right)$
that will be confirmed at time $k+i$ with a generalized Poisson model~\cite{chen_generalized_2016} 
with parameter $\lambda_\text{O}\left(k+i \mid k\right)$. 

Introducing $L_{\text{0},k+i}\left(k+j\mid k\right)$, the random
variable representing the number of parcels contributing to the load
at time $k+j$ among those to be ordered in the time slot $k+i$,
$i=1,\dots,j$, one may write the random variable describing the parcels
ordered after the time interval $k$ as 
\begin{equation}
L_{0}\left(k+j\mid k\right)=L_{\text{0},k+1}\left(k+j\mid k\right)+\dots+L_{\text{0},k+j}\left(k+j\mid k\right).\label{eq:load_future_do}
\end{equation}
One has 
\begin{align}
 \mathbb{P}\left(L_{\text{0},k+i}\left(k+j\mid k\right)=\ell\right)
 =\sum_{m=0}^{\infty}\mathbb{P}\left(L_{\text{0},k+i}\left(k+j\mid k\right)=\ell\mid O\left(k+i\mid k\right)=m\right)
 \cdot\mathbb{P}\left(O\left(k+i\mid k\right)=m\right).\label{eq:Poisson_do}
\end{align}
The pmf $\mathbb{P}\left(L_{\text{0},k+i}\left(k+j\mid k\right)=\ell\mid O\left(k+i\mid k\right)=m\right)$
is then evaluated as 
\begin{align}
 \mathbb{P}\left(L_{\text{0},k+i}\left(k+j\mid k\right)=\ell\mid O\left(k+i\mid k\right)=m\right)
 =\mathbb{P}\left({\textstyle \sum_{\tau\in\mathbb{V}_{k+i,m}}\Lambda_{\tau}\left(k+j\right)=\ell}\right),\label{eq:V_do}
\end{align}
where $\mathbb{V}_{k+i,m}$ is the set of virtual parcel indexes (since
not known at time $k$), with $\left|\mathbb{V}_{k+i,m}\right|=m$,
and such that for all $\tau\in\mathbb{V}_{k+i,m}$, $T_{0}\left(\tau\right)=k+i$.
Assuming again that the random variables $\Lambda_{\tau}\left(k+j\right)$,
$\tau\in\mathbb{V}_{k+i,m}$ are independent, the pmf of $\sum_{\tau\in\mathbb{V}_{k+i,m}^{r}}\Lambda_{\tau}\left(k+j\right)$
is obtained as the convolutions of the $m$ pmfs of $\Lambda_{\tau}\left(k+j\right)$
when $\tau\in\mathbb{V}_{k+i,m}$, which is evaluated using Proposition~\ref{prop:L0}.
Note that for all $\tau\in\mathbb{V}_{k+i,m}$, $r$ and $c$ are
unknown, as the parcel has not been ordered at time $k$. One may
choose $r$ and $c$ at random according to estimated retailer and
carrier selection probabilities $\mathbb{P}\left(R=r\right)$ and
$\mathbb{P}\left(C=c\mid R=r\right)$, as the carrier may depend on
the retailer.

\subsection{Load prediction algorithm}

\label{subsec:Load-prediction-algorithm}

Algorithm~\ref{Alg:LPA} summarizes the evaluation of the probability mass function of the load $p_{L,k+j\mid k}$ for time $k+j$ using knowledge available up to time $k$. From Line~2 to~7, the contribution to the load of each parcel in $\mathcal{P}_k^\rho$, \emph{i.e.}, which status is known at time $k$ is evaluated. The \texttt{load\_contrib} function provides the pmf of $\Lambda_\tau(k+j)$ using Proposition~1, 2, or~3, depending on the current known status $n$ of the parcel $\tau$. From Line~8 to~18, the contribution of parcels that will be ordered after time $k$ is evaluated. Each possible future order time instant $k+i$ is considered. Then, at Line~10, the pmf $\Lambda_{\tau}(k+i)$ is evaluated using Proposition~4. Only $m_{\max}$ terms in the sum \eqref{eq:Poisson_do} are considered. This requires a prediction of the expected number $\lambda_\text{O}(k+i\mid k)$ of future orders at Line~11. Then \eqref{eq:V_do} and 
\eqref{eq:Poisson_do} are evaluated iteratively at Line~16 and~17. Finally, at Line~18, the contribution of future orders is integrated in that of parcels which status is known at time $k$.

The functions \texttt{load\_contrib} and \texttt{future\_load\_contrib} invole the transition probability functions \eqref{eq:def_f_n}.
The way they may be estimated as well as the evaluation of $\lambda_\text{O}(k+i\mid k)$ is detailed in Section~\ref{Sec:Application}.  

\begin{algorithm}[h!]
\DontPrintSemicolon
\LinesNumbered
\SetAlFnt{\small}
\caption{Load prediction algorithm}
\label{Alg:LPA}
\SetKwInOut{Input}{Input}\SetKwInOut{Output}{Output}

\Input{$k$; $j$; $\mathcal{P}_k^\rho$; $\lambda_\text{O}$;} 
\Output{$p_{L,k+j\mid k}$}  

\Begin{
    \tcp{Init. pmf of $L(k+j \mid k)$}
    
    $p_{L,k+j\mid k} \longleftarrow [1,0,\dots,0]$

    \ForEach{$\tau\in\mathcal{P}_k^\rho$}{
       
			 \tcp{Eval. contrib. of $\tau$ to $L(k+j \mid k)$}

       $n \longleftarrow S_{k}\left(\tau\right)$ \tcp{Current status}
       
       $t_{n} \longleftarrow T_{n}\left(\tau\right)$ \tcp{Entering time}

       \tcp{Eval. $p_\tau$ pmf of $\Lambda_\tau(k+j)$ using Props 1-3 depending on $n$}
       $p_\tau \longleftarrow \FuncSty{load\_contrib}(n,t_n,k,j)$ 

       \tcp{Update pmf of $L(k+j \mid k)$}
        $p_{L,k+j\mid k} \longleftarrow \FuncSty{convolve}(p_{L,k+j\mid k}, p_\tau)$
           
    }
    \tcp{Init. pmf $p_\text{F}$ of $L_0(k+j\mid k)$ (contrib. of future orders)}
    
    $p_\text{F} \longleftarrow [1,0,\dots,0]$

    \tcp{Loop on future order times $k+i$}
    \ForEach{$i \in [1, j-1]$}{
            \tcp{Eval $p_\tau$ pmf of $\Lambda_{\tau}(k+i)$ see Prop. 4}

            $p_\tau \longleftarrow \FuncSty{future\_load\_contrib}(i,k)$ 

            \tcc{Eval. $\lambda_{\text{O}}(k+i\mid k)$ for Poisson model for nb of orders at time $k+i$}
            $\lambda \longleftarrow  \FuncSty{eval\_lambda(i,k)}$ 

             \tcc{$m_{max}$ s.t. Poisson cdf with param. $\lambda(k+i\mid k)$ is larger than 0.99}
            $m_\text{max} \longleftarrow  \FuncSty{ppf\_poisson}(\lambda(k+i\mid k),0.99)$ 

            \tcc{Init. pmf of $L_{\text{0},k+i}\left(k+j\mid k\right)$ knowing that $O\left(k+i\mid k\right)$}
            $p_0\longleftarrow[1,0,\dots,0]$ 

            \tcp{Init. pmf of $L_{\text{0},k+i}\left(k+j\mid k\right)$}
            $q_i \longleftarrow [1,0,\dots,0]$
            
            \tcp{For each nb $m$ of orders}
            
            \ForEach{$m \in [1,m_{\max}]$}{

                \tcp{Iter. eval. of \eqref{eq:V_do} using $p_m$}
                
                $p_m \longleftarrow \FuncSty{convolve}(p_{m-1},p_\tau)$ 

                \tcp{Iter. eval. of \eqref{eq:Poisson_do} using $q_i$}
                $q_i \longleftarrow q_i+p_m\cdot \frac{e^{-\lambda}\lambda^{m}}{m!}$ 
            }
        $p_\text{F} \longleftarrow \FuncSty{convolve}(p_\text{F},q_i)$
    }

     \tcp{Update $p_{L,k+j\mid k}$ with contrib. of orders after time $k$}
    $p_{L,k+j\mid k} \longleftarrow \FuncSty{convolve}(p_{L,k+j\mid k},p_\text{F})$

    \Return $p_{L,k+j\mid k}$\

}
\end{algorithm}

\section{Application\label{Sec:Application}}

\begin{figure}[ht]
\centering\includegraphics[width=0.5\columnwidth]{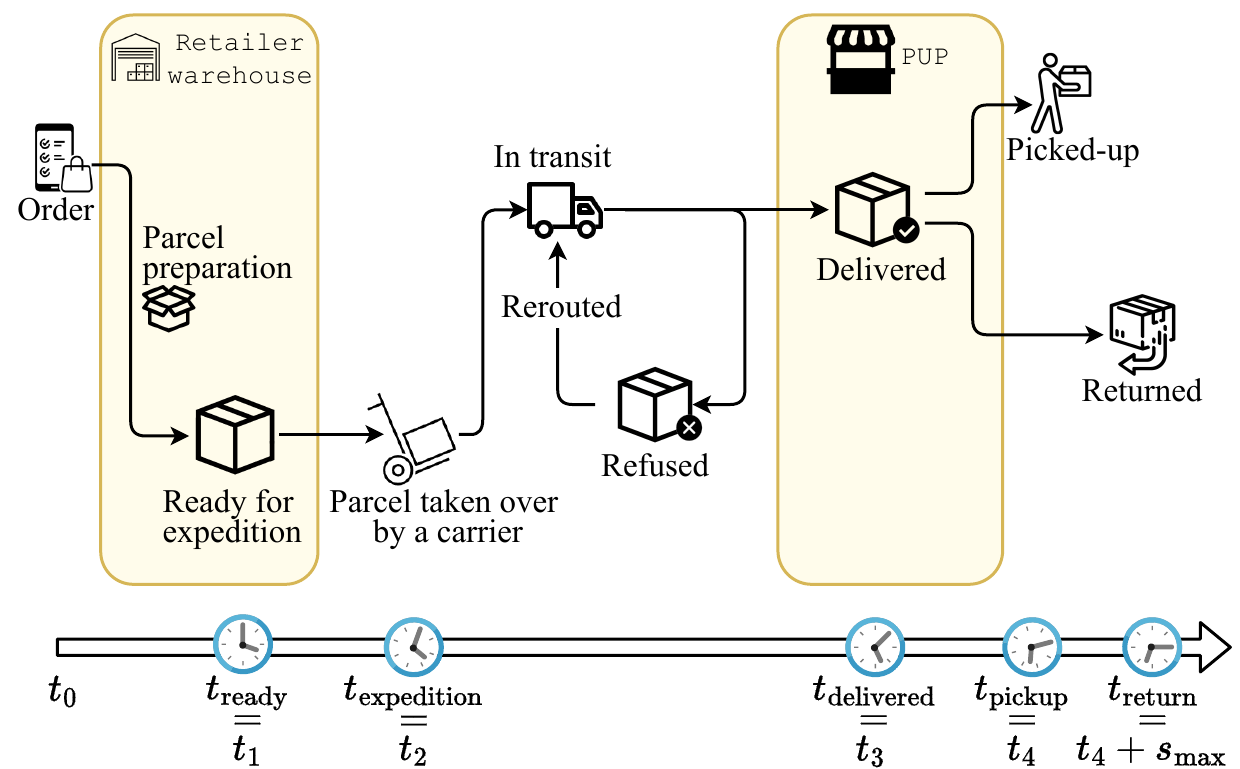}\caption{Life-cycle of a parcel\label{fig:schema }}
\end{figure}

Figure~\ref{fig:schema } shows the typical life-cycle of a parcel
$\tau$ containing a product purchased by a customer from an online
retailer at time $t_{0}$ and chosen to be delivered at a PUP $\rho$.
The order is processed by the retailer and is ready for expedition
at time $t_{\text{1}}.$ Carrier $C\left(\tau\right)\in\Gamma$ takes
the parcel over from the retailer warehouse at time $t_{\text{2}}$
and delivers it to the chosen PUP at time $t_{\text{3}}$. The delay
between $t_{\text{2}}$ and $t_{\text{3}}$ depends on $C\left(\tau\right)$
and on the relative location of the warehouse and of the PUP. Processing
at intermediate dispatch centers are integrated in the delay between
$t_{2}$and $t_{3}$. In most of the cases, the parcel is accepted
by the PUP and waits until it is picked up by the customer at time
$t_{\text{4}}$. The parcel is returned to the retailer warehouse
when its maximum sojourn time is reached. Some parcels may be refused
by the PUP in case of overload or closure and are rerouted to an alternative
PUP. This possibility is not considered in what follows.

\begin{figure}
\centering\includegraphics[width=0.5\columnwidth]{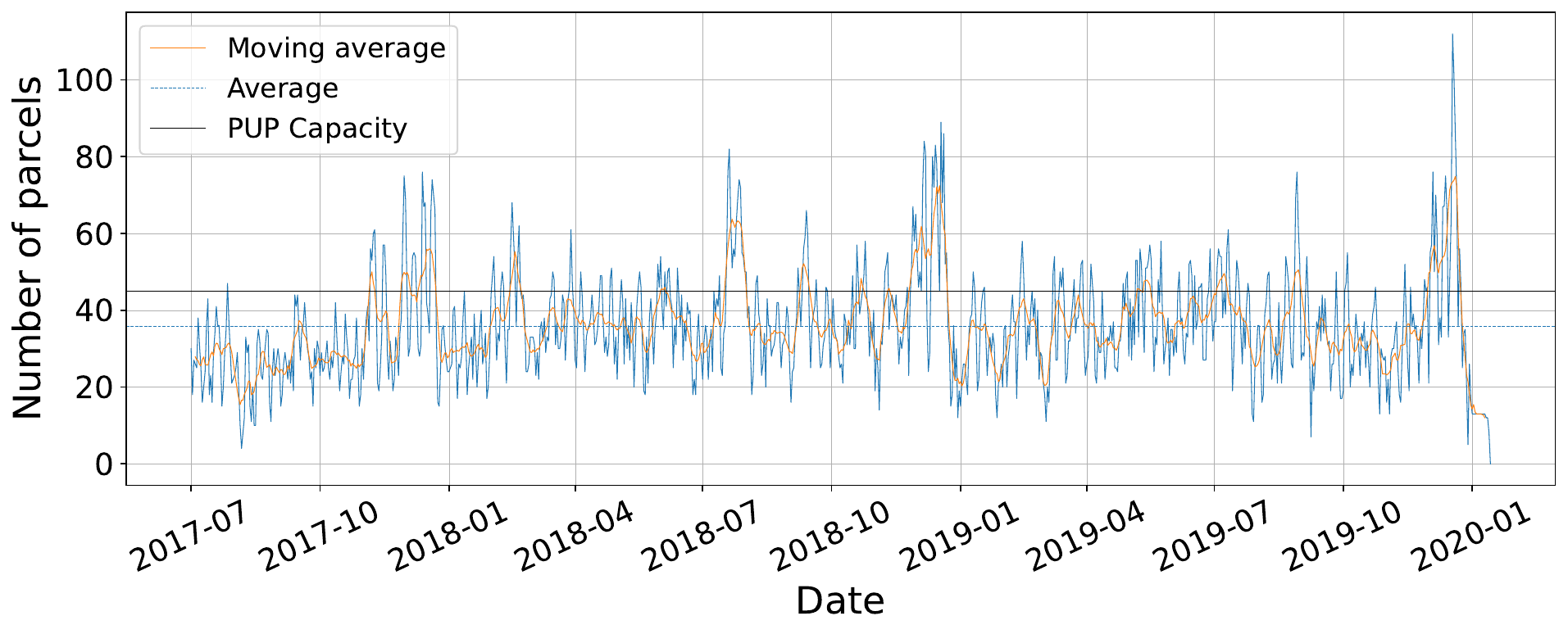}

\caption{Evolution of the load at 13:00 of the considered PUP from July 2017
to December 2019.\label{fig:Evolution-load}}
\end{figure}

To illustrate the proposed PUP load prediction approach, we consider
a local shop serving as PUP in Roussillon (France). Figure~\ref{fig:Evolution-load}
shows the evolution of its load at 13:00 from July 2017 to December
2019. The PUP has a capacity of $45$ parcels and is open from 9:00
to 19:00, from Monday to Saturday and from 9:00 to 12:00 on Sundays.
Sunday is not a working day for carriers in that area. This PUP has
had no long closing periods from 2017 to 2019. The data used to obtain
the load evolution are available in the database\footnote{https://github.com/cabani/ForecastingParcels}.

\subsection{Model}

In the considered database,
the time of order validation is not available. Moreover, only the
day at which parcels are ready to be delivered is available. Consequently,
in the life cycle of a parcel introduced in Section~\ref{Sec:Application},
only three statuses are considered, namely taken over ($S\left(\tau\right)=2$),
delivered ($S\left(\tau\right)=3$), and picked-up ($S\left(\tau\right)=4$).
The picked-up status is also associated to returned parcels, since
the effect of picked-up and returned parcels on the PUP load is the
same. Moreover, there are three possible carriers $\Gamma=\left\{ 1,2,3\right\} $.

For a parcel $\tau$ with status $S\left(\tau\right)=2$, the delay
between taken-over and delivery to the PUP $H_{2}\left(\tau\right)=T_{3}\left(\tau\right)-T_{2}\left(\tau\right)$
depends on the carrier $C=c$ (as the delivery delay differs among
carriers), and on the day of the week $w\left(T_{2}\left(\tau\right)\right)$
the parcel has been taken over, \emph{i.e.}, 
\[
\mathbb{P}\left(H_{3}=t_{3}-t_{2}\mid T_{2}=t_{2},C=c\right)=f_{2}\left(t_{3}-t_{2}\mid w\left(t_{2}\right),c\right).
\]
The delay between delivery and pick-up $H_{3}\left(\tau\right)=T_{4}\left(\tau\right)-T_{3}\left(\tau\right)$
depends on the day of the week and on the hour of delivery. Consequently,
the distribution of $H_{3}$ knowing $T_{3}=t_{3}$ is such that 
\begin{align}
\mathbb{P}\left(H_{3}=t_{4}-t_{3}\mid T_{3}=t_{3}\right) & =f_{3}\left(t_{4}-t_{3}\mid w\left(t_{3}\right),h\left(t_{3}\right)\right),\label{eq:f_3}
\end{align}
\emph{i.e.}, only the day of the week and the hour of $t_{3}$ are
accounted for in the variability of the pmf $f_{3}$ of $H_{3}\left(\tau\right)$.

\subsection{Estimation of the pmf \textmd{\normalsize{}$f_{n}$\label{subsec:Estimation-of-fn}}}

In this paper, empirical frequencies are evaluated to obtain the components
of the pmf $f_{n}$. Alternatively, parametric models could have been
considered.

As $f_{2}$ only depends on the day of the week $w$ a parcel has
been taken over and of the carrier $c$, the estimation of $f_{2}\left(\delta t\mid w,c\right)$
is performed considering $w=1,\dots,7$ and $c=1,2,3$. Consequently
only $3\times7$~different pmfs are estimated. For each of these
pmfs, delivery delays ranging from $\delta t=0$~h to $\delta t=100$~h,
are considered, since the delivery delay is generally within 5 days
(day off included). There is a relatively important variability in
the delivery hour among PUPs, only data related to parcels with the
PUP~$\rho$ as target are considered for the estimation of $f_{2}\left(\delta t\mid w,c\right)$.
Assuming that the estimation is performed at time~$k$, for given
values of $w$ and $c$, one gets
\[
\widehat{f}_{2}\left(\delta t\mid w,c\right)=\frac{\left|\left\{ \tau\in\mathcal{P}_{k}^{\rho}\mid h_{2}\left(\tau\right)=\delta t,w\left(t_{2}\left(\tau\right)\right)=w,t_{3}\left(\tau\right)\leqslant k,C\left(\tau\right)=c\right\} \right|}{\left|\left\{ \tau\in\mathcal{P}_{k}^{\rho}\mid w\left(t_{2}\left(\tau\right)\right)=w,C\left(\tau\right)=c,t_{3}\left(\tau\right)\leqslant k\right\} \right|},
\]
where $\mathcal{P}_{k}^{\rho}$ is the set of parcels for
which information is available at time $k$ with PUP~$\rho$ as target,
whatever their status. Moreover, $h_{2}\left(\tau\right)$, $t_{2}\left(\tau\right)$,
and $t_{3}\left(\tau\right)$ are the observed value of $H_{2}\left(\tau\right)$,
$T_{2}\left(\tau\right)$, and $T_{3}\left(\tau\right)$ for parcel
$\tau$.

The pmf $f_{3}$ of the delay between parcel delivery and pick-up
depends on the day of the week $w$ and hour $h$ of delivery. The
estimation of $f_{3}\left(\delta t\mid w,h\right)$ is thus performed
for $w=1,\dots,7$ and $h=h_{\text{o}}\left(w\right),\dots,h_{\text{c}}\left(w\right)$,
where $h_{\text{o}}\left(w\right)$ and $h_{\text{c}}\left(w\right)$
are the opening and closing hours of the PUP~$\rho$ for the week
day $w$. For the PUP~$\rho$, this represents about $76$~different
pmfs to estimate, each with $\delta t=0,\dots,336$~h, to account
for a maximum parcel sojourn time of two weeks in the PUP before being
returned. Assuming again an estimation performed at time~$k$, for
a given value of $w$ and $h$, one gets
\[
\widehat{f}_{3}\left(\delta t\mid w,h\right)=\frac{\left|\left\{ \tau\in\mathcal{P}_{k}^{\rho}\mid h_{3}\left(\tau\right)=\delta t,w\left(t_{3}\left(\tau\right)\right)=w,h\left(t_{3}\left(\tau\right)\right)=h,t_{4}\left(\tau\right)\leqslant k\right\} \right|}{\left|\left\{ \tau\in\mathcal{P}_{k}^{\rho}\mid w\left(t_{3}\left(\tau\right)\right)=w,h\left(t_{3}\left(\tau\right)\right)=h,t_{4}\left(\tau\right)\leqslant k\right\} \right|}
\]
where $h_{3}\left(\tau\right)$, $t_{3}\left(\tau\right)$,
and $t_{4}\left(\tau\right)$ are the observed value of $H_{3}\left(\tau\right)$,
$T_{3}\left(\tau\right)$, and $T_{4}\left(\tau\right)$ for parcel
$\tau$.

The pmfs $f_{n}$, $n=2,3$ account well for regular closing days,
\emph{e.g.}, on Sundays via the conditioning on the weekday $w$.
Nevertheless, on days off (New Year, First of May), there may be no
parcel collection or delivery. When the PUP is closed on days off
and for holidays, there are no parcel delivery and pick up too. PUP
closing days are usually known in advance by the PMC. The impact of
known closing days may be easily taken into account to update the
estimates of $f_{n}$.

\subsection{Prediction of the number of parcel orders\label{subsec:Prediction_DO}}

In the considered context, no information is available about parcel
orders. Consequently, instead of trying to estimate the number of
parcel that will be ordered at time $k+j$, one estimates the number
of parcels that will be taken over at time $k+i$, $i=1,\dots,j$
and which may contribute to the load of the PUP at time $k+j$.

Let $O\left(k\mid c\right)=\text{Card}\left(\left\{ \tau\,|\,T_{0}\left(\tau\right)=k,C\left(\tau\right)=c\right\} \right)$
be a random variable describing the number of parcels taken over at
time $k$ by Carrier $c$ and to be delivered to the PUP~$\rho$.
The sequence $\left\{ O\left(k\mid r\right)\right\} $ is a count
time series, described by a generalized Poisson model with time-varying
parameter $\lambda_{\text{O}}\left(k,c\right)$.

Considering the hourly breakdown of parcels taken over by Carrier
$c$ (with PUP $\rho$ as target) during a day, one has 
\begin{equation}
\lambda_{\text{O}}\left(k,r\right)=\rho_{w\left(k\right)}^{c}\left(h\left(k\right)\right)\mu_{c}\left(k-h\left(k\right)\right),\label{eq:hour_brd}
\end{equation}
where $\rho_{w}^{c}\left(h\right)\geqslant0$ is the proportion of
parcels taken over by Carrier $c$ during the time interval $\left[hT,\left(h+1\right)T\right[$
of the week day $w$ and is such that $\sum_{h}\rho_{w}\left(h\right)=1$,
$w=1,\dots,7$. Moreover, $\mu_{c}\left(k-h\left(k\right)\right)$
is the number of parcels taken over by Carrier $c$ during the whole
day starting at time $\left(k-h\left(k\right)\right)T$.

Using information available at time $k$, the parameters $\rho_{w}^{c}\left(h\right)$
are estimated as 
\begin{align}
\widehat{\rho}_{w}^{c}\left(h\right) & =\frac{\left|\left\{ \tau\in\mathcal{P}_{k}^{\rho}\mid w\!\left(T_{1}\!\left(\tau\!\right)\!\right)=w,h\!\left(T_{1}\!\left(\tau\!\right)\!\right)=h,C\!\left(\tau\!\right)=c\!\right\} \right|}{\left|\left\{ \tau\in\mathcal{P}_{k}\mid w\!\left(T_{1}\!\left(\tau\!\right)\!\right)=w,C\!\left(\tau\!\right)=c\!\right\} \right|},\label{eq:EstimRho}
\end{align}
where $\mathcal{P}_{k}^{\rho}$ is the set of parcels for which information
is available at time $k$, whatever their status and their target
PUP. The sequence $\left\{ \mu_{r}\right\} $ is a non-negative count
time series, for which we consider a SARIMA model~\cite{pongdatu_seasonal_2018}
to estimate future values.

\section{Results\label{sec:Results}}

Table~\ref{tab:MAE-and-MAPE} shows the prediction error of $L\left(k+j\mid k\right)$
considering a prediction performed at midnight of each day. The obtained
results are compared with the direct prediction of the load at 13:00
of each day considered as a time series described by a Holt-Winters~\cite{holt_forecasting_2004}
model, a SARIMA model, a SARIMAX model with the prediction of the
number of delivered parcels as exogenous variable, a Random Forest~\cite{breiman_random_2001}
model, several LSTM~\cite{dash_sales_2020} models, and our previously
parcel flow-based approach \cite{nguyen_load_2022}.

An MAE of 4.47 parcels is obtained for $j=13$ (one-day ahead prediction)
and of $8.12$ parcels for $j=85$ (four-day ahead prediction). These
results outperform the direct prediction using the other models, especially
the SARIMAX model that integrates also prior knowledge on parcels
already taken over by a carrier, as well as for our previously proposed
parcel flow-based approach. This illustrates the benefits of taking
into account the various parcels statuses during the delivery process.

\begin{table}[ht]
\centering
\begin{tabular}{|c|c|c|c|c|c|c|c|c|}
\hline 
\multirow{1}{*}{Approach} & \multicolumn{2}{c|}{$j=13$} & \multicolumn{2}{c|}{$j=37$} & \multicolumn{2}{c|}{$j=61$} & \multicolumn{2}{c|}{$j=85$}\tabularnewline
\hline 
 & MAE  & MAPE  & MAE  & MAPE  & MAE  & MAPE  & MAE  & MAPE\tabularnewline
\hline 
Holt-Winters  & 6.74  & 18.4  & 8.48  & 24.3  & 9.68  & 27.5  & 10.12  & 28.3\tabularnewline
\hline 
SARIMA  & 6.42  & 17.9  & 7.65  & 22.4  & 8.44  & 24.9  & 8.7  & 26.2\tabularnewline
\hline 
SARIMAX  & 5.33  & 14.40  & 7.03  & 20.30  & 7.99  & 23.60  & 8.45  & 25.1\tabularnewline
\hline 
Random Forest  & 6.85  & 20.3  & 8.67  & 26.9  & 9.28  & 30.2  & 9.44  & 30.8\tabularnewline
\hline 
Stacked-LSTM  & 9.30  & 26.4  & 11.49  & 38.1  & 12.31  & 36.5  & 12.94  & 39.3\tabularnewline
\hline 
Bi-LSTM  & 9.62  & 28.0  & 11.93  & 35.3  & 13.30  & 36.2  & 13.65  & 35.5\tabularnewline
\hline 
Conv-LSTM  & 8.20  & 23.2  & 10.37  & 32.7  & 11.16  & 36.6  & 11.03  & 41.5\tabularnewline
\hline 
Parcel flow \cite{nguyen_load_2022}  & 5.65  & 14.1  & 7.77  & 19.8  & 9.31  & 24.1  & 10.17  & 26.8\tabularnewline
\hline 
Parcel life-cycle  & \textbf{4.47}  & \textbf{12.9}  & \textbf{6.06}  & \textbf{18.4}  & \textbf{7.21}  & \textbf{21.2}  & \textbf{8.12}  & \textbf{23.7}\tabularnewline
\hline 
\end{tabular}

\caption{MAE and MAPE of the prediction errors of $L\left(k+j\mid k\right)$
for the proposed and alternative approaches.\label{tab:MAE-and-MAPE}}
\end{table}

\begin{figure}[ht!]
\centering\includegraphics[width=0.5\columnwidth]{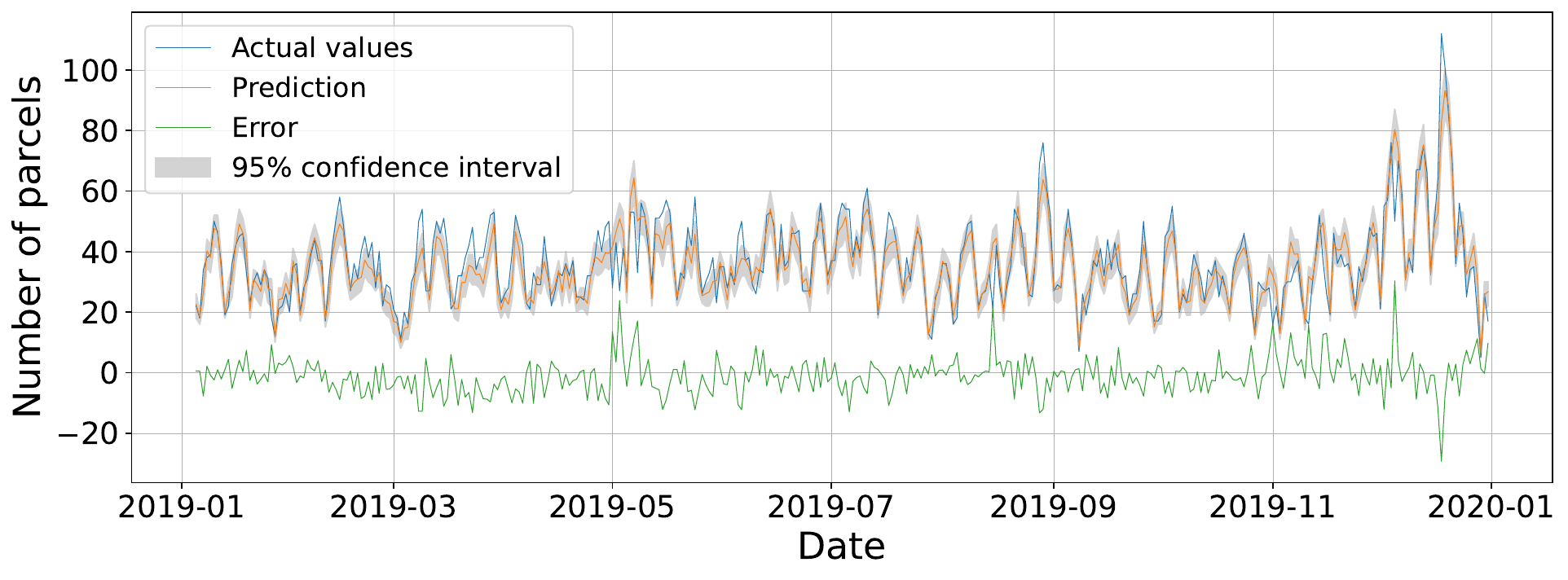}

\includegraphics[width=0.5\columnwidth]{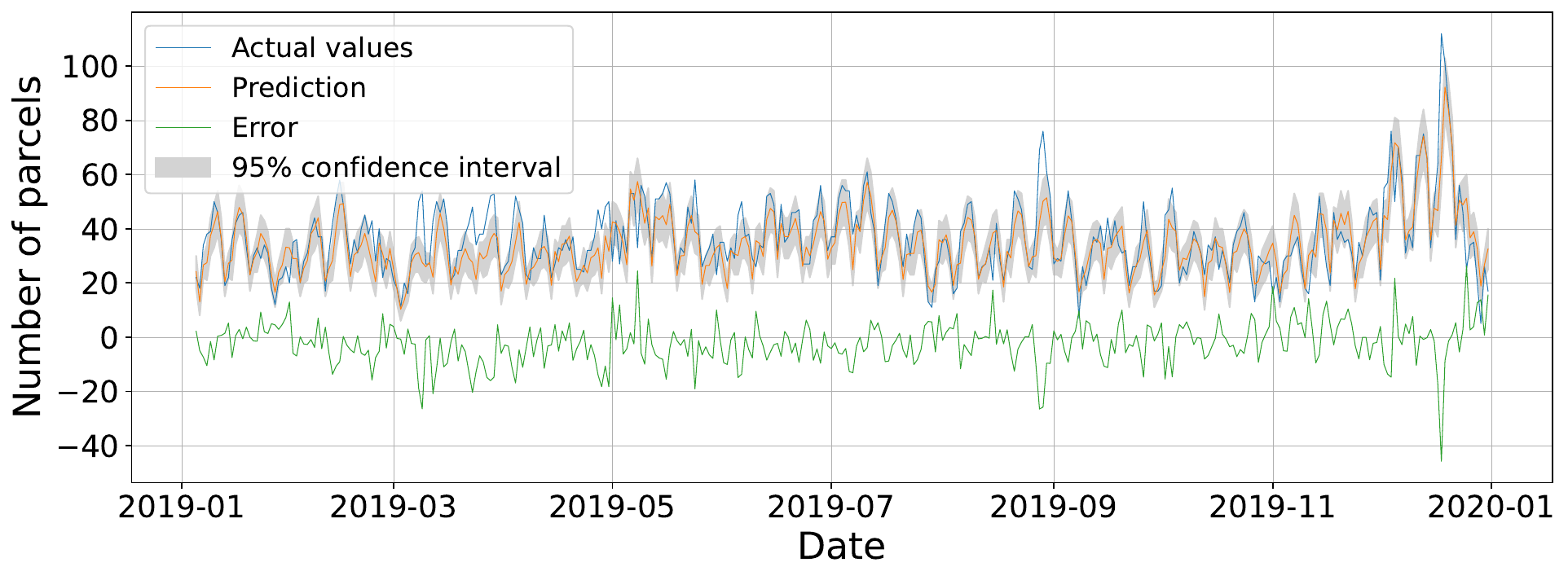}

\caption{Predicted and actual value of $L\left(k+13\right)$ (top) and $L\left(k+37\right)$
(bottom) and prediction errors\label{fig:Comparison-L(i)_1}}
\end{figure}

\begin{figure}[ht!]
\centering\includegraphics[width=0.5\columnwidth]{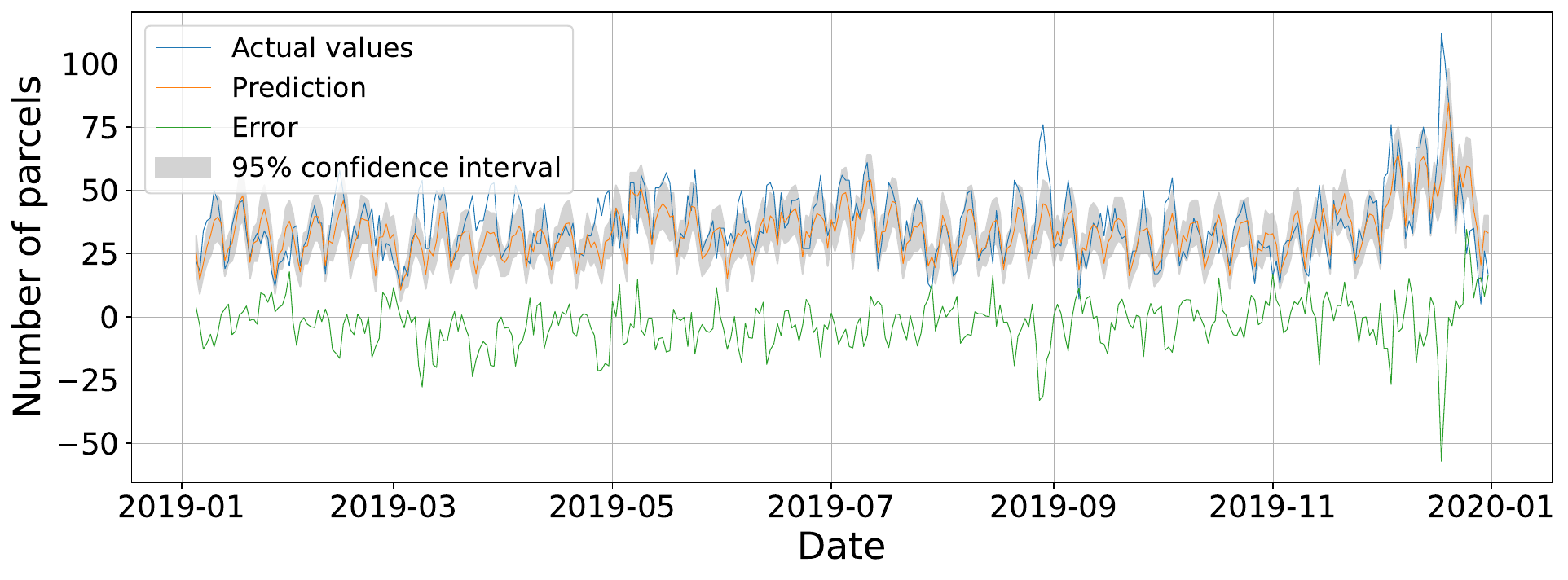}

\includegraphics[width=0.5\columnwidth]{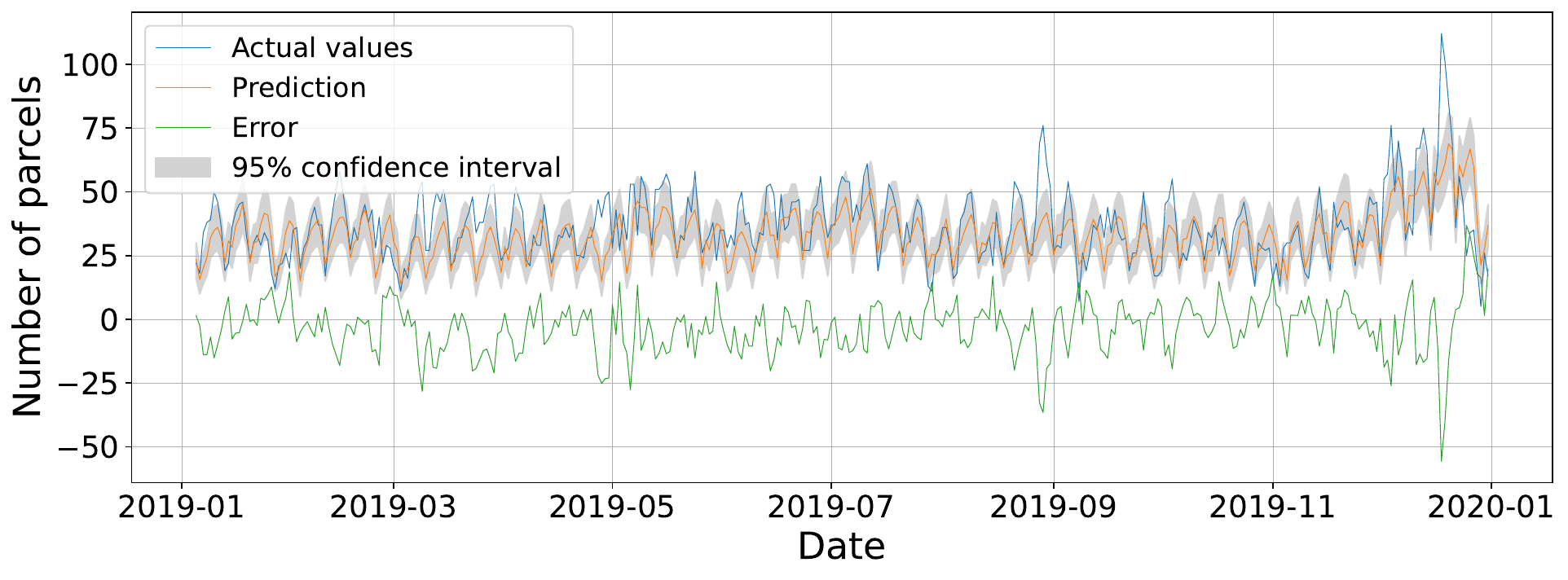}

\caption{Predicted and actual value of $L\left(k+61\right)$ (top) and $L\left(k+85\right)$
(bottom) and prediction errors\label{fig:Comparison-L(i)_4}}
\end{figure}

Figures~\ref{fig:Comparison-L(i)_1} to~\ref{fig:Comparison-L(i)_4}
compare the actual and predicted values of the load $L\left(k+j\mid k\right)$
for $k=13,37,61,85$. For $j\geqslant65$ (three and four day ahead
prediction), there is an important prediction error before Christmas.
The activity peak is not well predicted by the SARIMA model used in
the proposed approach to predict future orders.

\section{Conclusion\label{sec:Conclusion}}

This paper introduces a load forecasting approach for PUPs, leveraging
on a Markov jump process model of the life cycle of parcels. This
approach has been applied to predict the load related to B2C e-commerce.
The propose approach outperforms alternative techniques more agnostic
of the parcel life-cycle, which consider the load as a time series.

Future work includes increasing the prediction horizon to 7 days ahead
by gathering information on the number of future parcels to be processed
from the carriers and the online retailers. A load-balancing method
must be developed to distribute upcoming parcels among PUPs. The proposed
models must also be adapted to forecast the load of automatic parcel
lockers with more constraints due to the limited number of locker
units for parcels of multiple sizes.

\bibliographystyle{unsrt}

\appendix

\section{Proofs}

\subsection{Proof of Proposition~\ref{prop:Ld}}

\label{subsec:Proof_Prop_Ld}

At time $k$, consider a parcel $\tau$ with current status $S_{k}=N-1$,
\emph{i.e.}, that has been delivered at time $t_{N-1}\leqslant k$,
and that has not yet been picked up at time $k$. The probability
that $\tau\in\mathcal{L}_{N}\left(k+j|k\right)$, \emph{i.e.}, that
it is still in the PUP at time $k+j$, $j\geqslant1$ is 
\begin{equation}
\mathbb{P}\left(\Lambda_{\tau}\left(k+j\right)=1\mid T_{N-1}=t_{N-1},T_{N}>k\right)=\mathbb{P}\left(T_{N}>k+j\mid T_{N-1}=t_{N-1},T_{N}>k\right).\label{eq:L_d_1}
\end{equation}
Then (\ref{eq:L_d_1}) is rewritten as
\begin{align*}
\mathbb{P}\left(T_{N}>k+j\mid T_{N-1}=t_{N-1},T_{N}>k\right) & =\frac{\mathbb{P}\left(T_{N}>k+j,T_{N}>k\mid T_{N-1}=t_{N-1}\right)}{\mathbb{P}\left(T_{N}>k\mid T_{N-1}=t_{N-1}\right)}\\
 & =\frac{\mathbb{P}\left(T_{N}>k+j\mid T_{N-1}=t_{N-1}\right)}{\mathbb{P}\left(T_{N}>k\mid T_{N-1}=t_{N-1}\right)}\\
 & =\frac{1-\mathbb{P}\left(T_{N}\leqslant k+j\mid T_{N-1}=t_{N-1}\right)}{1-\mathbb{P}\left(T_{N}\leqslant k\mid T_{N-1}=t_{N-1}\right)}\\
 & =\frac{1-\sum_{t_{N}=t_{N-1}+1}^{k+j}f_{N-1}\left(t_{N}-t_{N-1}\mid t_{N-1}\right)}{1-\sum_{t_{N}=t_{N-1}+1}^{k}f_{N-1}\left(t_{N}-t_{N-1}\mid t_{N-1}\right)}
\end{align*}
where $f_{N-1}\left(t_{N}-t_{N-1}\mid t_{N-1}\right)$ is given by
\eqref{eq:def_f_n}.

\subsection{Proof of Proposition~\ref{prop:Lc}}

\label{subsec:Proof_Prop_Lc}

At time $k$, consider a parcel $\tau$ with current status $S_{k}=N-2$,
\emph{i.e.}, that has reached a dispatch center at time $t_{N-2}\leqslant k$
and has not been delivered at time $k$. The probability that $\tau\in\mathcal{L}_{N-1}\left(k+j|k\right)$,
\emph{i.e.}, that it is delivered before time $k+j$ and still in
the PUP at time $k+j$, $j\geqslant1$ is 
\begin{align}
\mathbb{P}\left(\Lambda_{\tau}\left(k+j\right)=1\mid T_{N-2}=t_{N-2},T_{N-1}>k\right) & =\mathbb{P}\left(T_{N-1}\leqslant k+j,T_{N}>k+j\mid T_{N-2}=t_{N-2},T_{N-1}>k\right),\label{eq:L_c_1}
\end{align}
One may rewrite (\ref{eq:L_c_1}) as 
\begin{align}
\mathbb{P}\left(\Lambda_{\tau}\left(k+j\right)=1\mid T_{N-2}=t_{N-2},T_{N-1}>k\right) & =\mathbb{P}\left(T_{N-1}\leqslant k+j\mid T_{N-2}=t_{N-2},T_{N-1}>k\right)\nonumber \\
 & \mathbb{P}\left(T_{N}>k+j\mid k<T_{N-1}\leqslant k+j,T_{N-2}=t_{N-2}\right).\label{eq:L_c_2}
\end{align}
Consider the first term of~(\ref{eq:L_c_2}), 
\begin{align}
\mathbb{P}\left(T_{N-1}\leqslant k+j\mid T_{N-2}=t_{N-2},T_{N-1}>k\right) & =\frac{\mathbb{P}\left(T_{N-1}\leqslant k+j,T_{N-1}>k\mid T_{N-2}=t_{N-2}\right)}{\mathbb{P}\left(T_{N-1}>k\mid T_{N-2}=t_{N-2}\right)}\nonumber \\
 & =\frac{\sum_{t_{N-1}=k+1}^{k+j}\mathbb{P}\left(T_{N-1}=t_{N-1}\mid T_{N-2}=t_{N-2}\right)}{1-\sum_{t_{N-1}=t_{N-2}+1}^{k}\mathbb{P}\left(T_{N-1}=t_{N-1}\mid T_{N-2}=t_{N-2}\right)}\nonumber \\
 & =\frac{\sum_{t_{N-1}=k+1}^{k+j}f_{N-2}\left(t_{N-1}-t_{N-2}\mid t_{N-2}\right)}{1-\sum_{t_{N-1}=t_{N-2}+1}^{k}f_{N-2}\left(t_{N-1}-t_{N-2}\mid t_{N-2}\right)}.\label{eq:L_c_3}
\end{align}
The second term of~(\ref{eq:L_c_2}) is 
\begin{align}
\mathbb{P}\left(T_{N}>k+j\mid k<T_{N-1}\leqslant k+j,T_{N-2}=t_{N-2}\right) & =1-\mathbb{P}\left(T_{N}\leqslant k+j\mid k<T_{N-1}\leqslant k+j,T_{N-2}=t_{N-2}\right)\nonumber \\
 & =1-\frac{\mathbb{P}\left(T_{N}\leqslant k+j,k<T_{N-1}\leqslant k+j\mid T_{N-2}=t_{N-2}\right)}{\mathbb{P}\left(k<T_{N-1}\leqslant k+j\mid T_{N-2}=t_{N-2}\right)}\nonumber \\
 & =1-\frac{\sum_{t_{N-1}=k+1}^{k+j-1}\mathbb{P}\left(T_{N}\leqslant k+j,T_{N-1}=t_{N-1}\mid T_{N-2}=t_{N-2}\right)}{\sum_{t_{N-1}=k+1}^{k+j}\mathbb{P}\left(T_{N-1}=t_{N-1}\mid T_{N-2}=t_{N-2}\right)}.\label{eq:L_c_4}
\end{align}
In (\ref{eq:L_c_4}), one has 
\begin{align}
\mathbb{P}\left(T_{N}\leqslant k+j,T_{N-1}=t_{N-1}\mid T_{N-2}=t_{N-2}\right) & =\mathbb{P}\left(T_{N}\leqslant k+j\mid T_{N-1}=t_{N-1},T_{N-2}=t_{N-2}\right)\nonumber \\
 & \cdot\mathbb{P}\left(T_{N-1}=t_{N-1}\mid T_{N-2}=t_{N-2}\right)\nonumber \\
 & =\mathbb{P}\left(T_{N}\leqslant k+j\mid T_{N-1}=t_{N-1}\right)\nonumber \\
 & \cdot f_{N-2}\left(t_{N-1}-t_{N-2}\mid t_{N-2}\right).\label{eq:L_c_5}
\end{align}
using the Markov property and \eqref{eq:def_f_n}. Then 
\begin{align}
\mathbb{P}\left(T_{N}\leqslant k+j\mid T_{N-1}=t_{N-1}\right) & =\sum_{t_{N}=k+1}^{k+j}\mathbb{P}\left(T_{N}=t_{N}\mid T_{N-1}=t_{N-1}\right)\nonumber \\
 & =\sum_{t_{N}=k+1}^{k+j}f_{N-1}\left(t_{N}-t_{N-1}\mid t_{N-1}\right).\label{eq:L_c_6}
\end{align}
Finally, combining (\ref{eq:L_c_2}) to (\ref{eq:L_c_6}), one gets
\begin{align}
\mathbb{P}\left(\Lambda_{\tau}\left(k+j\right)=1\mid T_{N-2}=t_{N-2},T_{N-1}>k\right) & =\frac{\sum_{t_{N-1}=k+1}^{k+j}f_{N-2}\left(t_{N-1}-t_{N-2}\mid t_{N-2}\right)}{1-\sum_{t_{N-1}=t_{N-2}+1}^{k}f_{N-2}\left(t_{N-1}-t_{N-2}\mid t_{N-2}\right)}\nonumber \\
 & \hspace{-3cm} \left(1-\frac{\sum_{t_{N-1}=k+1}^{k+j-1}\sum_{t_{N}=k+1}^{k+j}f_{N-1}\left(t_{N}-t_{N-1}\mid t_{N-1}\right)f_{N-2}\left(t_{N-1}-t_{N-2}\mid t_{N-2}\right)}{\sum_{t_{N-1}=k+1}^{k+j}f_{N-2}\left(t_{N-1}-t_{N-2}\mid t_{N-2}\right)}\right).\label{eq:L_c_7}
\end{align}

\subsection{Proof of Proposition~\ref{prop:Ln}}

\label{subsec:Proof_Prop_Ln}

At time~$k$, consider a parcel $\tau$ with current status $S_{k}=n$,
reached at time $t_{n}$, with $0<n<N-1$. This parcel is shipped
by carrier $C\left(\tau\right)=c$ to the target PUP $\rho$. The
probability that $\tau\in\mathcal{L}_{n}\left(k+j\mid k\right)$,
\emph{i.e.}, that it is delivered before time $k+j$ and still in
the PUP at time $k+j$, $j\geqslant1$ is
\begin{align}
\mathbb{P}\left(\Lambda_{\tau}\left(k+j\right)=1\mid T_{n}=t_{n},T_{n+1}>k\right) & =\mathbb{P}\left(T_{N-1}\!\leqslant k+j,T_{N}\!>k+j\mid T_{n}=t_{n},T_{n+1}\!>k\right).\label{eq:L_n_1}
\end{align}
Using the conditional probability, (\ref{eq:L_n_1}) can be rewritten
as 
\begin{align}
\mathbb{P}\left(\Lambda_{\tau}\left(k+j\right)=1\mid T_{n}=t_{n},T_{n+1}>k\right) & =\mathbb{P}\left(T_{N-1}\leqslant k+j\mid T_{n}=t_{n},T_{n+1}>k\right)\nonumber \\
 & \mathbb{P}\left(T_{N}>k+j\mid T_{n}=t_{n},T_{n+1}>k,T_{N-1}\leqslant k+j\right).\label{eq:L_n_2}
\end{align}

Consider the first term of~(\ref{eq:L_n_2}), 
\begin{align}
\mathbb{P}\left(T_{N-1}\leqslant k+j\mid T_{n}=t_{n},T_{n+1}>k\right) & =\frac{\mathbb{P}\left(T_{N-1}\leqslant k+j,T_{n+1}>k\mid T_{n}=t_{n}\right)}{\mathbb{P}\left(T_{n+1}>k\mid T_{n}=t_{n}\right)}\nonumber \\
 & =\frac{\sum_{t_{N-1}=k+1}^{k+j}\sum_{t_{n+1}=k+1}^{t_{N-1}}\mathbb{P}\left(T_{N-1}=t_{N-1},T_{n+1}=t_{n+1}\mid T_{n}=t_{n}\right)}{1-\sum_{t_{n+1}=t_{n}+1}^{k}\mathbb{P}\left(T_{n+1}=t_{n+1}\mid T_{n}=t_{n}\right)}\label{eq:L_n_3}
\end{align}
In the denominator of \eqref{eq:L_n_3}, one has
\[
\mathbb{P}\left(T_{n+1}=t_{n+1}\mid T_{n}=t_{n}\right)=\sum_{t_{n+1}=t_{n}+1}^{k}f_{n}\left(t_{n+1}-t_{n}\mid t_{n}\right)
\]
For the numerator of \eqref{eq:L_n_3}, one has
\begin{align*}
\mathbb{P}\left(T_{N-1}=t_{N-1},T_{n+1}=t_{n+1}\mid T_{n}=t_{n}\right) & =\mathbb{P}\left(T_{N-1}=t_{N-1}\mid T_{n+1}=t_{n+1},T_{n}=t_{n}\right)\\
 & \cdot\mathbb{P}\left(T_{n+1}=t_{n+1}\mid T_{n}=t_{n}\right)\\
 & =\mathbb{P}\left(T_{N-1}=t_{N-1}\mid T_{n+1}=t_{n+1}\right)\\
 & \cdot\mathbb{P}\left(T_{n+1}=t_{n+1}\mid T_{n}=t_{n}\right).
\end{align*}
Then 
\begin{align*}
\mathbb{P}\left(T_{N-1}=t_{N-1}\mid T_{n+1}=t_{n+1}\right) & =\sum_{t_{n+2}=t_{n+1}+1}^{t_{N-1}}\mathbb{P}\left(T_{N-1}=t_{N-1},T_{n+2}=t_{n+2}\mid T_{n+1}=t_{n+1}\right)\\
 & =\sum_{t_{n+2}=t_{n+1}+1}^{t_{N-1}}\mathbb{P}\left(T_{N-1}=t_{N-1}\mid T_{n+2}=t_{n+2}\right)\\
 & \cdot\mathbb{P}\left(T_{n+2}=t_{n+2}\mid T_{n+1}=t_{n+1}\right).
\end{align*}
This process may be iterated up to $\mathbb{P}\left(T_{N-1}=t_{N-1}\mid T_{N-2}=t_{N-2}\right)$
to get finally 
\begin{align*}
\mathbb{P}\left(T_{N-1}\leqslant k+j,T_{n+1}>k\mid T_{n}=t_{n}\right) & =\sum_{t_{N-1}=k+1}^{k+j}\sum_{t_{n+1}=k+1}^{t_{N-1}}\sum_{t_{n+2}=t_{n+1}+1}^{t_{N-1}}\dots\sum_{t_{N-2}=t_{N-3}+1}^{t_{N-1}}\mathbb{P}\left(T_{N-1}=t_{N-1}\mid T_{N-2}=t_{N-2}\right)\dots\nonumber\\
&\mathbb{P}\left(T_{n+2}=t_{n+2}\mid T_{n+1}=t_{n+1}\right)\mathbb{P}\left(T_{n+1}=t_{n+1}\mid T_{n}=t_{n}\right)\\
 & =\sum_{t_{N-1}=k+1}^{k+j}g_{n}^{N-1}\left(t_{n},t_{N-1}\right)
\end{align*}
where 
\begin{align*}
g_{n}^{N-1}\left(t_{n},t_{N-1}\right)=&\sum_{t_{n+1}=k+1}^{t_{N-1}}\sum_{t_{n+2}=t_{n+1}+1}^{t_{N-1}}\dots\sum_{t_{N-2}=t_{N-3}+1}^{t_{N-1}}f_{N-2}\left(t_{N-1}-t_{N-2}\mid t_{N-2}\right)\nonumber\\
&\dots f_{n+1}\left(t_{n+2}-t_{n+1}\mid t_{n+1}\right)f_{n}\left(t_{n+1}-t_{n}\mid t_{n}\right)
\end{align*}
which is the probability for a parcel to switch from status $n$ in
which it is at time $t_{n}$ to status $N-1$ at time $t_{N-1}$.
Consequently, the first term of~(\ref{eq:L_n_2}) is
\begin{align}
\mathbb{P}\left(T_{N-1}\leqslant k+j\mid T_{n}=t_{n},T_{n+1}>k\right) & =\frac{\sum_{t_{N-1}=k+1}^{k+j}g_{n}^{N-1}\left(t_{n},t_{N-1}\right)}{1-\sum_{t_{n+1}=t_{n}+1}^{k}f_{n}\left(t_{n+1}-t_{n}\mid t_{n}\right)}.\label{eq:L_n_First_term}
\end{align}

Consider the second term of (\ref{eq:L_n_2}),
\begin{align}
\mathbb{P}\left(T_{N}>k+j\mid T_{n}=t_{n},T_{n+1}>k,T_{N-1}\leqslant k+j\right) & =1-\mathbb{P}\left(T_{N}\leqslant k+j\mid T_{n}=t_{n},T_{n+1}>k,T_{N-1}\leqslant k+j\right)\nonumber \\
 & =1-\frac{\mathbb{P}\left(T_{N}\leqslant k+j,T_{N-1}\leqslant k+j,T_{n+1}>k\mid T_{n}=t_{n}\right)}{\mathbb{P}\left(T_{N-1}\leqslant k+j,T_{n+1}>k\mid T_{n}=t_{n}\right)}.\label{eq:L_n_4}
\end{align}
Consider the denominator of (\ref{eq:L_n_4}),
\begin{align*}
\mathbb{P}\left(T_{N-1}\leqslant k+j,T_{n+1}>k\mid T_{n}=t_{n}\right) & =\sum_{t_{N-1}=k+1}^{k+j}\sum_{t_{n+1}=k+1}^{t_{N-1}}\mathbb{P}\left(T_{N-1}=t_{N-1},T_{n+1}=t_{n+1}\mid T_{n}=t_{n}\right)\\
 & =\sum_{t_{N-1}=k+1}^{k+j}\sum_{t_{n+1}=k+1}^{t_{N-1}}\sum_{t_{n+2}=t_{n+1}+1}^{t_{N-1}}\mathbb{P}\left(T_{N-1}=t_{N-1},T_{n+2}=t_{n+2},T_{n+1}=t_{n+1}\mid T_{n}=t_{n}\right)\\
 & =\sum_{t_{N-1}=k+1}^{k+j}\sum_{t_{n+1}=k+1}^{t_{N-1}}\sum_{t_{n+2}=t_{n+1}+1}^{t_{N-1}}\mathbb{P}\left(T_{N-1}=t_{N-1},T_{n+2}=t_{n+2}\mid T_{n+1}=t_{n+1}\right)\\
 & \mathbb{P}\left(T_{n+1}=t_{n+1}\mid T_{n}=t_{n}\right).
\end{align*}
This process may again be iterated to get
\begin{align}
\mathbb{P}\left(T_{N-1}\leqslant k+j,T_{n+1}>k\mid T_{n}=t_{n}\right) & =\sum_{t_{N-1}=k+1}^{k+j}\sum_{t_{n+1}=k+1}^{t_{N-1}}g_{n+1}^{N-1}\left(t_{n+1},t_{N-1}\right)\mathbb{P}\left(T_{n+1}=t_{n+1}\mid T_{n}=t_{n}\right)\nonumber \\
 & =\sum_{t_{N-1}=k+1}^{k+j}\sum_{t_{n+1}=k+1}^{t_{N-1}}g_{n+1}^{N-1}\left(t_{n+1},t_{N-1}\right)f_{n}\left(t_{n+1}-t_{n}\mid t_{n}\right).\label{eq:L_n_secondterm_den}
\end{align}
Consider the numerator of (\ref{eq:L_n_4})
\begin{align}
\mathbb{P}&\left(T_{N}\leqslant k+j,T_{N-1}\leqslant k+j,T_{n+1}>k\mid T_{n}=t_{n}\right)=\nonumber\\
				&\sum_{t_{N}=k+1}^{k+j}\sum_{t_{N-1}=k+1}^{t_{N}}\sum_{t_{n+1}=k+1}^{t_{N-1}}\mathbb{P}\left(T_{N}=t_{N},T_{N-1}=t_{N-1},T_{n+1}=t_{n+1}\mid T_{n}=t_{n}\right)
\end{align}
Then
\begin{align*}
\mathbb{P}\left(T_{N}=t_{N},T_{N-1}=t_{N-1},T_{n+1}=t_{n+1}\mid T_{n}=t_{n}\right) & =\mathbb{P}\left(T_{N}=t_{N}\mid T_{N-1}=t_{N-1},T_{n+1}=t_{n+1},T_{n}=t_{n}\right)\\
 & \cdot\mathbb{P}\left(T_{N-1}=t_{N-1},T_{n+1}=t_{n+1}\mid T_{n}=t_{n}\right)\\
 & =\mathbb{P}\left(T_{N}=t_{N}\mid T_{N-1}=t_{N-1}\right)\\
 & \cdot\mathbb{P}\left(T_{N-1}=t_{N-1},T_{n+1}=t_{n+1}\mid T_{n}=t_{n}\right)\\
 & =f_{N-1}\left(t_{N}-t_{N-1}\mid t_{N-1}\right)g_{n+1}^{N-1}\left(t_{n+1},t_{N-1}\right)f_{n}\left(t_{n+1}-t_{n}\mid t_{n}\right).
\end{align*}
Then
\begin{align}
\mathbb{P}&\left(T_{N}\leqslant k+j,T_{N-1}\leqslant k+j,T_{n+1}>k\mid T_{n}=t_{n}\right)=\nonumber\\
& \sum_{t_{N}=k+1}^{k+j}\sum_{t_{N-1}=k+1}^{t_{N}}\sum_{t_{n+1}=k+1}^{t_{N-1}}f_{N-1}\left(t_{N}-t_{N-1}\mid t_{N-1}\right)g_{n+1}^{N-1}\left(t_{n+1},t_{N-1}\right)f_{n}\left(t_{n+1}-t_{n}\mid t_{n}\right)\label{eq:L_n_secondterm_num}
\end{align}

Finally, combining \eqref{eq:L_n_First_term}, \eqref{eq:L_n_secondterm_den},
and \eqref{eq:L_n_secondterm_num}, one gets
\begin{align*}
 & \mathbb{P}\left(\Lambda_{\tau}\left(k+j\right)=1\mid T_{n}=t_{n},T_{n+1}>k\right)=\frac{\sum_{t_{N-1}=k+1}^{k+j}g_{n}^{N-1}\left(t_{n},t_{N-1}\right)}{1-\sum_{t_{n+1}=t_{n}+1}^{k}f_{n}\left(t_{n+1}-t_{n}\mid t_{n}\right)}\\
 & \cdot\left(1-\frac{\sum_{t_{N}=k+1}^{k+j}\sum_{t_{N-1}=k+1}^{t_{N}}\sum_{t_{n+1}=k+1}^{t_{N-1}}f_{N-1}\left(t_{N}-t_{N-1}\mid t_{N-1}\right)g_{n+1}^{N-1}\left(t_{n+1},t_{N-1}\right)f_{n}\left(t_{n+1}-t_{n}\mid t_{n}\right)}{\sum_{t_{N-1}=k+1}^{k+j}\sum_{t_{n+1}=k+1}^{t_{N-1}}g_{n+1}^{N-1}\left(t_{n+1},t_{N-1}\right)f_{n}\left(t_{n+1}-t_{n}\mid t_{n}\right)}\right).
\end{align*}

\subsection{Proof of Proposition~\ref{prop:L0}}
\label{App:Proof:L0}

At time $k$, consider a parcel $\tau$ that is expected to be ordered
at time $T_{0}\left(\tau\right)=t_{0}$ with $k<t_{0}<k+j$ with target
PUP $\rho$. The parcel preparation duration $H_{0}\left(\tau\right)=T_{1}\left(\tau\right)-T_{0}\left(\tau\right)$
depends on the retailer $r$, and the delivery time depends on the
carrier $c$. In the rest of the proof we omit explicit references
to the carrier $c$ and the retailer $r$. The probability that $\tau\in\mathcal{L}_{\text{0}}\left(k+j\mid k\right)$,
\emph{i.e.}, that it is delivered before time $k+j$ and still in
the PUP at time $k+j$, $j\geqslant1$ is
\begin{align}
\mathbb{P}\left(L_{0}\left(k+j\right)=1\mid T_{0}=t_{0}\right) & =\mathbb{P}\left(T_{N-1}\leqslant k+j,T_{N}>k+j\mid T_{0}=t_{0}\right).\label{eq:L_0_1}
\end{align}

Using the properties of conditional probability, (\ref{eq:L_0_1})
can be rewritten as 
\begin{align}
\mathbb{P}\left(L_{0}\left(k+j\right)=1\mid T_{0}=t_{0}\right) & =\mathbb{P}\left(T_{N-1}\leqslant k+j\mid T_{0}=t_{0}\right)\nonumber \\
 & \times\mathbb{P}\left(T_{N}>k+j\mid T_{0}=t_{0},T_{N-1}\leqslant k+j\right).\label{eq:L_0_2}
\end{align}
Consider the first factor of~(\ref{eq:L_0_2}), 
\begin{align}
\mathbb{P}\left(T_{N-1}\leqslant k+j\mid T_{0}=t_{0}\right) & =\sum_{t_{N-1}=k+1}^{k+j}\mathbb{P}\left(T_{N-1}=t_{N-1}\mid T_{0}=t_{0}\right)\label{eq:L_0_3}
\end{align}
Moreover one has
\begin{align*}
\mathbb{P}\left(T_{N-1}=t_{N-1}\mid T_{0}=t_{0}\right) & =g_{0}^{N-1}\left(t_{0},t_{N-1}\right).
\end{align*}
Consequently,
\begin{equation}
\mathbb{P}\left(T_{N-1}\leqslant k+j\mid T_{0}=t_{0}\right)=\sum_{t_{N-1}=k+1}^{k+j}g_{0}^{N-1}\left(t_{0},t_{N-1}\right).\label{eq:L_0:First_term}
\end{equation}

Consider the second factor of (\ref{eq:L_0_2}),
\begin{align}
\mathbb{P}\left(T_{N}>k+j\mid T_{0}=t_{0},T_{N-1}\leqslant k+j\right) & =1-\mathbb{P}\left(T_{N}\leqslant k+j\mid T_{0}=t_{0},T_{N-1}\leqslant k+j\right)\nonumber \\
 & =1-\frac{\mathbb{P}\left(T_{N}\leqslant k+j,T_{N-1}\leqslant k+j\mid T_{0}=t_{0}\right)}{\mathbb{P}\left(T_{N-1}\leqslant k+j\mid T_{0}=t_{0}\right)}.\label{eq:L_0_4}
\end{align}
The denominator of (\ref{eq:L_0_4}) simplifies
\begin{align}
\mathbb{P}\left(T_{N-1}\leqslant k+j\mid T=t_{0}\right) & =\sum_{t_{N-1}=t_{0}+N-1}^{k+j}\mathbb{P}\left(T_{N-1}=t_{N-1}\mid T_{0}=t_{0}\right)\nonumber \\
 & =\sum_{t_{N-1}=t_{0}+N-1}^{k+j}g_{0}^{N-1}\left(t_{0},t_{N-1}\right).\label{eq:L_0:Second_Term}
\end{align}
Consider the numerator of (\ref{eq:L_0_4})
\begin{equation}
\mathbb{P}\left(T_{N}\leqslant k+j,T_{N-1}\leqslant k+j\mid T_{0}=t_{0}\right)=\sum_{t_{N-1}=t_{0}+N-1}^{k+j-1}\sum_{t_{N}=t_{N-1}+1}^{k+j}\mathbb{P}\left(T_{N}=t_{N},T_{N-1}=t_{N-1}\mid T_{0}=t_{0}\right),\label{eq:L_0:Third_term}
\end{equation}
in which each term of the double sum evaluates to
\begin{align}
\mathbb{P}\left(T_{N}=t_{N},T_{N-1}=t_{N-1}\mid T_{0}=t_{0}\right) & =\mathbb{P}\left(T_{N}=t_{N}\mid T_{N-1}=t_{N-1},T_{0}=t_{0}\right)\mathbb{P}\left(T_{N-1}=t_{N-1}\mid T_{0}=t_{0}\right)\nonumber \\
 & =\mathbb{P}\left(T_{N}=t_{N}\mid T_{N-1}=t_{N-1}\right)\mathbb{P}\left(T_{N-1}=t_{N-1}\mid T_{0}=t_{0}\right)\nonumber \\
 & =f_{N-1}\left(t_{N}-t_{N-1}\mid t_{N-1}\right)g_{0}^{N-1}\left(t_{0},t_{N-1}\right).\label{eq:L_0:Last_term}
\end{align}

Finally, combining \eqref{eq:L_0_2}, \eqref{eq:L_0:Second_Term},
\eqref{eq:L_0:Third_term}, and \eqref{eq:L_0:Last_term}, one gets
\begin{align*}
\mathbb{P}\left(\Lambda_{\tau}\left(k+j\right)=1\mid T_{0}=t_{0}\right)= & \sum_{t_{N-1}=k+1}^{k+j}g_{0}^{N-1}\left(t_{0},t_{N-1}\right)\\
 & \cdot\left(1-\frac{\sum_{t_{N-1}=t_{0}+N-1}^{k+j-1}\sum_{t_{N}=t_{N-1}+1}^{k+j}f_{N-1}\left(t_{N}-t_{N-1}\mid t_{N-1}\right)g_{0}^{N-1}\left(t_{0},t_{N-1}\right)}{\sum_{t_{N-1}=t_{0}+N-1}^{k+j}g_{0}^{N-1}\left(t_{0},t_{N-1}\right)}\right).
\end{align*}

\end{document}